%% file: Paper.tex
\documentclass[iop,apj,numberedappendix]{emulateapj}
 \usepackage{graphicx}
  \usepackage{apjfonts}
  \usepackage{amssymb} 
  \usepackage{amsmath}
\slugcomment{Accepted for publication in ApJ, Feb 1, 2013  }
 \shorttitle{Protoplanetary disk structure with grain evolution: \\ the ANDES model}
 \shortauthors{Akimkin et al.}

\begin{document}
 %%%%%%%%%%%%%%%%%%%%%%%%%%%%%%%%%%%%%%%%%%%%%%%%%%%%%%%%%%%%%%%%% 
 \newcommand{\HH }{{\ensuremath{\rm H_2\,}}} 
 \newcommand{\mum}{\ensuremath{\rm \mu m\,}}
 \newcommand{\Ms}{\ensuremath{\rm M_{\sun}}} 
 \newcommand{\Msyr}{\ensuremath{\rm  M_{\sun}\,yr^{-1}}} 
 \newcommand{\Mspc}{\ensuremath{\rm M_{\sun}\,pc^{-2}}}
 \newcommand{\ddt}[1]{{{\rm d}\, {#1} \over{\rm d}\,t}}
 \newcommand{\gcmc}{\ensuremath{\,\rm g\,cm^{-3}}} 
 \newcommand{\cmc}{\ensuremath{\,\rm  cm^{-3}}} 
 \newcommand{\cmq}{\ensuremath{\,\rm cm^{-2}}}
 \newcommand{\Zsun}{\ensuremath{\rm Z_{\sun}}}
 \newcommand{\dd}[2]{{{\rm d}\, {#1} \over{\rm d}\,#2}}
 %----------------------------------------------------------------------------

 \title{Protoplanetary disk structure with grain evolution: \\ The ANDES model}

 \author{V.~Akimkin$^1$, S.~Zhukovska$^2$, D.~Wiebe$^1$, D.~Semenov$^2$, Ya.~Pavlyuchenkov$^1$,
   A.~Vasyunin$^3$, T.~Birnstiel$^4$, Th.~Henning$^2$} 
 \affil{$^1$Institute of Astronomy of the RAS, Pyatnitskaya str. 48, Moscow, Russia}
 \affil{$^2$Max-Planck-Institute for Astronomy, K\"onigstuhl 17, D-69117 Heidelberg,
 Germany} 
 \affil{$^3$Department of Chemistry, the University of Virginia, USA }
 \affil{$^4$Harvard-Smithsonian Center for Astrophysics, 60 Garden Street, Cambridge, MA 02138, USA 
 \and    Excellence Cluster Universe, Technische Universit\"at M\"unchen, Boltzmannstr. 2, 85748 Garching, Germany} 
% \email{akimkin, dwiebe, pavyar@inasan.ru, zhukovska, semenov, henning@mpia.de, anton.vasyunin@gmail.com, tbirnstiel@cfa.harvard.edu}
 \email{akimkin@inasan.ru}
 \begin{abstract} 

 We present a  self-consistent model of a protoplanetary
 disk: `ANDES' (`AccretioN disk with Dust Evolution and Sedimentation'). ANDES is based on a flexible and extendable modular structure that includes 1)~a  1+1D~frequency-dependent continuum radiative transfer module, 2)~a module to calculate the chemical evolution using an extended gas-grain
 network with UV/X-ray-driven processes surface reactions, 3)~a module to calculate the gas thermal energy
 balance, and 4)~a~1+1D module that simulates dust grain evolution. For the first time, grain evolution  and
time-dependent molecular chemistry are included in a protoplanetary disk model.
We find that grain growth and sedimentation of large grains to the disk midplane lead to a
 dust-depleted atmosphere.  Consequently, dust and gas temperatures  become higher in the inner disk ($R\la 50$~AU)
and lower in the outer disk ($R\ga 50$~AU), in comparison with the disk model with pristine dust.
 The response of disk chemical structure to the dust growth and
 sedimentation is twofold. First, due to higher transparency a partly UV-shielded
 molecular layer is shifted closer to the dense midplane. Second, the presence of big
 grains in the disk midplane delays the freeze-out of volatile gas-phase species such as
 CO there, while in adjacent upper layers the depletion is still effective.
Molecular concentrations and thus column
densities of many species are enhanced in the disk model with dust evolution, e.g.,
CO$_2$, NH$_2$CN, HNO, H$_2$O, HCOOH, HCN, CO.
We also show that time-dependent chemistry is important for a proper description of gas thermal
 balance. 
\end{abstract}

 \keywords{accretion, accretion disks -- circumstellar matter -- stars: formation --
 stars: pre-main-sequence, astrochemistry}

 \input intro.tex

 \input model.tex

 \input results_and_discussions.tex

 \input conclusions.tex 
 \input acknowledgements.tex
%\footnotesize \bibliography{ref}

\input{ref.bbl}
 \input appendix.tex

 \end{document}

%% file: intro.tex
\section{Introduction} 
The planet formation and, in particular, the origin of the Solar
System are among the most fascinating astrophysical problems that are far from being fully understood.
The quickly growing number of detected exoplanets hints to
ubiquitous planet formation in our Galaxy.
Space-born facilities (e.g., {\it Hubble, Spitzer,
Herschel}) as well as ground-based observatories (e.g., VLT, Keck, Subaru, PdBI, IRAM 30-m, SMA, early ALMA)
provide unique information on the appearance, structure, chemical composition, and
evolution of nearby protoplanetary disks
\citep[e.g.,][]{1997A&A...317L..55D, 2004ApJ...605L..53F, 2005ApJ...631.1134A, 2007ApJ...662.1067H, 2007prpl.conf..767N, 
2010pdac.book...97S, 2010A&A...518L.129S, Muto_ea12}. Relatively compact sizes of $\sim 100-1\,000$~AU and
low masses of $\sim0.01$~M$_{\odot}$ make disks a challenging target for observational studies.

Another obstacle to investigate the formation of planets is an enormous
range of physical conditions encountered in a protoplanetary disk and a wide variety of
interrelated processes 
\citep[e.g.,][]{2011arXiv1103.0556W}. The combined action of these processes defines the
appearance of the disk in scattered light, dust continuum, and atomic and molecular lines.
Modeling of continuum and line radiation implies
knowing stellar spectrum, dust density, dust temperature, and  size distribution as well
as gas density, gas temperature, and molecular content throughout the disk, and
in full 3D. If all this information is available, a multi-dimensional radiation transfer
(RT) model can be used to build a synthetic disk map at any wavelength
\citep[e.g.,][]{1992ApJ...395..529W,MH_97,1999A&A...349..839W,Dullemond:2002}.
Due to computational difficulties to follow global disk
evolution in 3D-MHD, particularly, coupled with  chemical kinetics models,
and the lack of necessary constraints related to the
magnetic field structure, turbulence, grain size distribution, etc., a disk model
needs to be simplified. One can steadily approach the warranted level of
physical complexity by adding new components to the model (e.g., going from 1D to 2D
geometry or from gray to non-gray radiative transfer) and comparing with observations at each
development step.

A number of disk models has been developed over time \citep[see review
in][]{2007prpl.conf..555D}. These models have been based on an RT-based disk structure (either
1D, 1+1D/2D,  or 3D), molecular abundances, and dust and gas thermal balance.
Disk models with detailed vertical structure and thermal balance
regulated solely by dust heating and cooling, and, in some cases, accretion heating,
have been developed by, e.g.,
\citet{1997ApJ...486..372B,1997ApJ...490..368C,MH_97,2004A&A...417..159D,
2005A&A...442..703H}. It has been typically assumed in such studies that the dust is
well mixed with the gas, and its properties do not differ from properties of the ISM dust. One
of the most widely used models of this kind has been developed by
\citet{1998ApJ...500..411D,1999ApJ...527..893D}. It has been extensively used in many
subsequent studies as a template of the disk density and temperature distribution
\citep[e.g.,][]{2001ApJ...547.1077C,2004A&A...417...93S,2006ApJS..165..568F}. Other
similar models, utilizing more accurate frequency-dependent RT algorithms or other
improvements (e.g., a full 2D geometry, evolving disk structure, more realistic dust
opacities) have been presented by
\citet{2001A&A...379..515M,2002A&A...395..853D,2002ApJ...567..587N,2009ApJ...705.1237G},
to name a few.

An important development of the protoplanetary disk models was to account for the energy
balance of dust and gas separately in dilute disk regions. There the rate of gas-dust
collisions drops so low that the gas becomes  thermally decoupled from the dust
\citep[e.g.,][]{2004A&A...428..511J,2004ApJ...615..991K,2008ApJ...683..287G}. The most recent and most advanced addition to
this family, the ``ProDiMo'' model, is presented by \citet{2009A&A...501..383W} and
updated in \citet{2011MNRAS.412..711T} and \citet{2011A&A...526A.163A}. This model is
based on iterative calculations of a 1+1D vertical hydrostatic disk structure, 2D
frequency-dependent dust continuum RT, gas-grain and FUV-photochemistry to calculate
abundances of molecular coolants, and an escape probability method to model non-LTE
heating and cooling of the gas. It is derived from thermo-chemical models of
\citet{2000A&A...353..276K}, \citet{2001A&A...373..641K}, and \citet{2004ApJ...615..991K}. Since 2011 it
includes X-ray-driven chemistry and heating via H$_2$ ionization and Coulomb heating
\citep{2011A&A...526A.163A}. Uniform dust abundances and power-law size distributions are typically
assumed  \citep{2012A&A...547A..69A}, with opacities for a dust mixture calculated by Effective Medium Theory
\citep{1935AnP...416..636B}. Abundances of molecules are calculated assuming chemical
equilibrium and element conservation, which may not be a valid approach to disk chemical
evolution  \citep[e.g.,][]{1976ARA&A..14...81B,IHMM04,2011ApJS..196...25S}.

 Recent observations at IR and mm-/cm-wavelengths  have shown that many disks around
young stars of ages $\ga 1$~Myr  have already a deficit of of small grains in the inner regions, $r\la
10-50$~AU and the presence of large, pebble-sized dust grains in the midplanes  compared to the pristine
ISM dust \citep[e.g.,][]{2011arXiv1103.0556W,2012M&PS..tmp..232W}. From the analysis of SEDs at millimeter and
centimeter wavelengths, grain sizes of at least $1$~cm have been inferred for many disks
 \citep[e.g.,][]{Rodmann_ea06,Lommen_ea09,Lommen_ea10a,Ricci_ea10a,2011ApJ...739L...7M,2012ApJ...760L..17P}.
 \citet{Guilloteau_ea11a} have used high-resolution interferometric PdBI
observations to discern dust emissivity slopes at millimeter wavelengths in a sample of
young stars. Their analysis has shown that in the Taurus-Auriga star-forming region some
disks show very low dust emissivity indices in the inner regions, characteristic of grains
with sizes of $\ga1~$mm, and slopes that are indicative of smaller grains toward the
disk edges. In addition, Spitzer IR spectroscopy of silicate bands at 10 and 20~$\mu$m has
revealed efficient crystallization and growth of the sub-micron-sized ISM grains in warm disk atmospheres in
many young systems, regardless of their ages, accretion rates, and disk masses
\citep[e.g.,][]{2006ApJ...639..275K,2009ApJ...703.1964F,2010ApJ...721..431J,2010ApJS..188...75M,
2011ApJ...734...51O,2011arXiv1108.5258S}. The dust settling associated with grain growth reduces disk
scale heights and flaring angles, and thus leads to less intense mid-IR disk emission than
expected from conventional hydrostatic models with uniform dust, in accordance with
observations of most T~Tauri stars \citep{2011arXiv1103.0556W}.

As dust is a very important ingredient of the disk physics,  evolution of its properties
should also be considered in disk models.  Usually both the grain growth and
sedimentation are accounted for in disk models in a parameterized
way, by assuming an increased upper limit of grain size $a_{\rm max}$ and artificially changing
the dust density  and the slope of dust size distribution in various disk regions.
For example, expanding on their earlier works,
\citet{2001ApJ...553..321D} have studied the influence of dust evolution on the disk
structure and its spectral energy distribution (SED). Grain growth has been simulated as
an increase of $a_{\rm max}$ up to 10~cm and change of the dust size distribution slope
$p$ from $-3.5$ to $-2.5$. In these models dust has been assumed to be well-mixed with the gas.

To study the effect of dust
settling, \citet{2006ApJ...638..314D}  have included two dust populations in the model,  with different spatial
distributions. \citet{2006ApJ...638..314D} shown that the evolved dust model better reproduces
observed millimeter fluxes and spectral slopes. A similar approach  to study the effect of dust settling on
the disk thermal and chemical structure has been taken by
\citet{2004A&A...428..511J} and \citet{Fogel_ea11}. Settling  has been simulated using variable dust/gas
mass ratio. A variable $a_{\rm max}$ value  has been used by \citet{2006ApJ...642.1152A} to
investigate changes in disk density, gas and dust temperature, and molecular abundances
due to dust growth.

More accurate methods to model dust growth are mainly based on solving the coagulation
(Smoluchowski) equation. Here the main attribute of the model is whether the dust
evolution is computed for a fixed disk structure or the dust evolution and disk structure
are mutually consistent. The first approach is used, e.g., in 
\citet{2006ApJ...640.1099N,2004ApJ...614..960S,2005ApJ...625..414T,2007ApJ...654L.159C}, who used
parameterized disk structure. The second approach
has been used by \citet{1997A&A...325..569S,2005ApJ...625..414T,2007ApJ...661..334N,2007ApJ...661..374T}.

An efficient scheme to tackle the modeling of dust coagulation, fragmentation,
sedimentation, turbulent stirring  around a `snow line' in a protoplanetary disk has been proposed by
\citet{2008A&A...487L...1B}.  They have found that major factors affecting grain
evolution are trapping of dust particles in gas pressure maxima and the
presence of a turbulently quiescent `dead zone' in disk inner midplane.
\citet{Birnstiel:2010p9709} have updated this model by considering
time-dependent  viscous evolution of a gas disk. They have found that  dust properties, gas pressure
gradients, and the strength of turbulence are more important factors for dust evolution than the initial conditions and the
early formation phase of the protoplanetary disk. \citet{2011A&A...525A..11B} have shown that, upon evolution, grain size
distribution reaches a quasi-steady state, which however, does not follow the standard
MRN-like power-law size distribution and is sensitive to the gas surface density, amount
of turbulence, and disk thermal structure.

The next step in protoplanetary disk modeling was made by~\citet{Vasyunin2011}, where detailed dust 
evolution was considered along with comprehensive set of gas-phase and surface
chemical reactions. However, to calculate disk thermal structure, they take into account only two heating sources, namely, viscous
dissipation and dust grain irradiation by the central star. It was shown that column densities of some molecules (like C$_2$H, HC$_{2n+1}$N ($n =$ 0--3), H$_2$O and C$_2$H$_2$/HCN abundance ratio) can be used as observational tracers of early stages of the grain evolution in protoplanetary disks.

 In this paper, for the first time, we consider the influence of dust evolution on
the disk structure by combining the detailed computation of the radiation field with 
the dust growth, fragmentation, and sedimentation model. When computing the disk density 
and temperature we take into account the full grain size distribution as a function 
of location in the disk. Gas temperature and dust temperature are computed separately, 
with taking into account the disk chemical structure. These two factors represent
a major improvement in comparison with \citet{Vasyunin2011}'s model. Also, a new 
detailed RT treatment is implemented with high frequency resolution from ultraviolet 
to far infrared. The organization of the paper is the following. In Section~2 the disk 
model ``ANDES'' (AccretioN Disk with Dust Evolution and Sedimentation) is described. 
In Section~3 we present a physical structure for a typical protoplanetary disk computed 
both for both pristine dust and for evolved dust. Also, the chemical structure is 
described in this section, and specific features of the disk chemical compositions are 
presented for various dust models. Discussion and conclusions follow. Details of 
gas energy balance processes and benchmarking results are presented in Appendix~\ref{AppendA} 
and~\ref{AppendB}.

%% file: model.tex
\section{Disk physical structure} 
A multitude of processes (gas dynamics, dust evolution,
energy transport processes, chemistry, etc.) makes modeling of protoplanetary disks a
challenge. With the current level of computing resources a  global 3D radiative MHD
simulation, including gas and dust evolution  and chemical kinetics, remains a topic for the future (but see, e.g., \citet{2012ApJ...761...95F} for such models). Nevertheless 
a sufficient understanding of protoplanetary disk
physics may be achieved by  detailed modeling of primary processes that govern its structure and
observational characteristics, and  simplified modeling of secondary processes. This makes the problem
tractable. For Class II objects \citep{Lada1987,Evans2009} it is usually assumed that a
disk  structure is in a steady-state regime  over a time span of $\sim1 $~Myr.
This is supported by observations of  disk kinematics via molecular lines and
 disk surface densities via (sub-)millimeter dust emissivity.
The   line profiles are indicative of Keplerian motion in  most of the disks
\citep{Koerner1993,Guilloteau1998,Pietu_ea07}. The estimates of the disk masses and density distributions show that
self-gravity is negligible for Class II objects \citep{Isella2009}. The assumption that these disks evolve on a
diffusion timescale and not on a hydrodynamical one allows setting aside hydrodynamical
simulations and reducing a 3D problem to a 1+1D problem. The azimuthal dimension is eliminated due
to the axial symmetry of an  unperturbed disk. The other two dimensions are usually split into
the radial structure that is determined by diffusive evolution, and the vertical structure that is derived from
the hydrostatic equilibrium equation \citep{1998ApJ...500..411D,Dullemond:2002}.
The 1+1D description is suitable for
dust continuum radiation transfer. For  disk regions outward of a few~AU a radial optical
depth for a location  close to the midplane is higher than the vertical  optical depth, so that the dust temperature is
mostly determined by vertical   diffusion of radiation. A gain in computation time that is
acquired by  a 1D radiation transfer, compared  to a 2D RT, allows better frequency
resolution, which is  important for dust temperature calculations
\citep{Dullemond:2002} and for  modeling photochemistry.

In this paper we adopt a 1+1D approach  to calculate disk density and temperature.
As the disk consist of two main ingredients~(dust and gas), the overall problem is reduced
to  calculating four physical quantities: dust and gas temperatures  ($T_{\rm d}, T_{\rm g}$), and dust and gas densities
($\rho_{\rm d}, \rho_{\rm g}$). This allows  to split the disk model
into four blocks, calculating the corresponding  quantities  at
each disk radius $R$:

\begin{itemize} 
\item[I.] \textbf{Dust temperature}. Dust temperature and radiation field $J_{\nu}$ are found by solving the
{\em radiation transfer\/} problem in vertical direction. The following quantities are considered
as input: the dust density, its optical properties (absorption and scattering
coefficients $\kappa_{\nu},\sigma_{\nu}$), external irradiation and all necessary
parameters describing non-radiative dust heating functions.

\item[II.] \textbf{Gas temperature}. To determine the gas temperature, we solve the local energy balance
equation, accounting for various heating and cooling processes. Since gas heating and
cooling rates depend on abundances of main heating/cooling species and their level
populations, it is necessary to  include \textit{chemical reactions} and simplified \textit{line radiation transport} in the gas
temperature calculation. 

\item[III.] \textbf{Dust density}. The \textit{dust evolution} is an essential part of our disk
model. The surface density of dust is assumed to be equal to 1\% of the  total gas density, whereas its
 detailed vertical structure and size distribution are determined from the dust growth and sedimentation physics.
 We consider coagulation and fragmentation of dust grains   and their redistribution due to turbulent stirring
and gravitational settling to the midplane. We also consider disk
structure with   for comparison. 

\item[IV.] \textbf{Gas density}. We assume that the gas vertical structure is defined by the local \textit{hydrostatic
equilibrium}. In this case the gas
density $\rho_{\rm g}$ can be found if its temperature $T_{\rm g}$, mean molecular weight
$\mu$, and surface density $\Sigma_g$ are known. The surface density is assumed
to be given by the predefined function $\Sigma_g(R)$.

\end{itemize}

\begin{figure}[!ht]
\begin{center}
\includegraphics[scale=0.3]{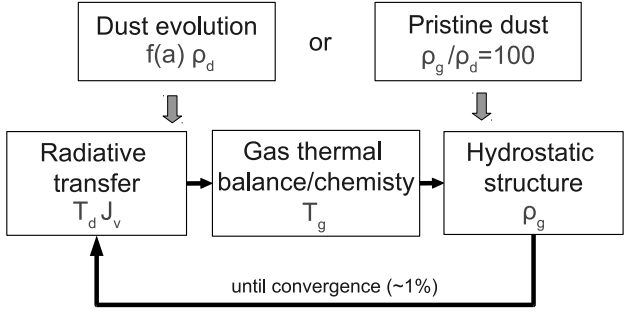}
\end{center} 
\caption{ Overall computational scheme for ANDES.} 
\label{flowchart} 
\end{figure}

As all these quantities are not independent, we iterate between the modules until
convergence is reached. The overall computational flowchart for ANDES is shown in
Figure~\ref{flowchart}. Assuming the surface density profile, we calculate dust evolution for 2~Myr starting from the MRN initial distribution. The resultant dust structure is then used to derive radiation field and gas disk structure using radiation transfer, energy balance and hydrostatic structure modules. As a fiducial dust model we also consider pristine grains with the following parameters: 0.1$\mu$m in size, astronomical silicate, dust to gas ratio 0.01 at every location in the disk. The list of basic assumptions is: (i) a disk is quasi-static, axially symmetric and treated in 1+1D approach; (ii) the gas surface density is assumed to be specified, and the dust surface density is 0.01 of the gas surface density; (iii) gas vertical structure is determined from the hydrostatic equilibrium, while dust vertical structure is a consequence of turbulent stirring and grain settling.  Also, calculating the chemical evolution we keep the dust properties fixed both for pristine and evolved dust cases.
Below we describe each part of the model in detail. 

\input RT.tex

\input GTB.tex 
\input HS.tex
\input DE.tex

%% file: RT.tex
\subsection{Radiative transfer}
The radiation in a protoplanetary disk plays a two-fold role. First, it is a main
energy carrier that redistributes energy coming from the stellar irradiation and viscous
dissipation, and thus defines the overall disk structure. Second, it determines rates of
photoreactions and  thus shapes the disk chemical structure and observational appearance. These two
aspects pose different requirements to the radiation transfer model. The radiation field
as a contributor to the disk energy balance should be known in a wide range of
wavelengths, from FUV (radiation from the accretion region  and non-thermal radiation from the central star)
to visual (thermal stellar radiation) to the infrared and submillimeter wavelengths (thermal disk radiation).
 This requirement makes multi-dimensional RT approaches with high spectral resolution too slow for
iterative disk modeling.

On the other hand,  in a narrow range of UV wavelengths (from 912\,\AA~to, say, 4000\,\AA)
 such a good spectral resolution is important for  accurate calculation of the photochemical rates, as
the dependence of  photoreaction cross-sections on $\lambda$ is  complicated.
Protoplanetary disks usually have high optical depths at $\lambda \la 100\,\mu$m \citep[e.g.,][]{1991ApJ...381..250B},
which calls for using suitable methods to solve the radiation transfer (RT)  problem for optically thick media.
As our primary focus is on the chemical  modeling in disks with evolved dust, we  developed such a method
 with a particularly good wavelength resolution in the UV part of the spectrum.

\subsubsection{Main equations} \label{MainEquationRT}
It is easy to show that in the cases of Schwarzschild-Schuster and Eddington approximations the RT equation for a plane-parallel
1D medium can be written using the mean intensity $J_{\nu}$:
\begin{equation}\label{MainEq} 
\frac{q}{\chi_{\nu}(z)} \frac{\partial}{\partial z}\left[\frac{1}{\chi_{\nu}(z)}\frac{\partial J_{\nu}(z)}{\partial z}\right]=J_{\nu}(z)-S_{\nu}(z), 
\end{equation} 
where $\chi_{\nu}$[\,cm$^{-1}$] is the extinction coefficient, $S_{\nu}$ is the
source function, and $q=1/4$ and $q=1/3$ for the
Schwarzschild-Schuster and the Eddington approximation,  respectively.

If we consider only dust continuum absorption, thermal emission, and coherent isotropic
scattering, the source function is 
\begin{equation}\label{S}
S_{\nu}(z)=\frac{\kappa_{\nu}(z)B_{\nu}\left(T_{\rm
d}\left(z\right)\right)+\sigma_{\nu}(z)J_{\nu}(z)}{\kappa_{\nu}(z)+\sigma_{\nu}(z)}.
\end{equation} 
Here $\kappa_{\nu}[{\rm cm}^{-1}$] is the absorption coefficient,
$\sigma_{\nu} [{\rm cm}^{-1}]$ is the scattering coefficient
($\chi_{\nu}=\kappa_{\nu}+\sigma_{\nu}$) and $B_{\nu}$ is the Planck function. 

In the 1+1D approach the anisotropic scattering by dust grains can also be taken into account. It is important
for UV photons interacting with small dust grains, whereas at IR wavelengths 
scattering can be considered negligible compared to absorption/emission. 
The $p$ parameter describing anisotropy of scattering ($p>1/2$ and
$p<1/2$ denote forward and backward scattering,
respectively) can be introduced in our RT model in such a
way that dust extinction efficiency $\sigma$ is substituted by the
combination $2(1-p)\sigma$. In the limit of
predominantly forward scattering grains, the role of UV dust
heating in deep disk layers
renders less significant than in the case of the isotropic scattering
used in our study. That is, our current approach tends to slightly overestimate the role of
scattering and thus overall dust heating in disk upper layers.

Equations~\eqref{MainEq}--\eqref{S} are closed with the energy balance equation
\begin{equation}\label{EEBFinal} 
4\pi\int\limits_{0}^{\infty}\kappa_{\nu}(z)
B_{\nu}(T_{\rm d}(z))\,d\nu = 4\pi\int\limits_{0}^{\infty}\kappa_{\nu}(z) J_{\nu}(z)\,d\nu
+\Gamma_{\rm nr}(z). 
\end{equation} 
Here $\Gamma_{\rm nr}(z)$ [\,erg cm$^{-3}$s$^{-1}]$ 
accounts for non-radiative heating/cooling mechanisms (gas-grain interaction, see Equation~\eqref{gasgrain}).

Equations~\eqref{MainEq}--\eqref{EEBFinal} represent the complete system for $J_{\nu}(z)$
and $T_{\rm d}(z)$. We solve this system with the analogue of the Feautrier method
\citep{Mihalas1978}. Specifically, we introduce a wavelength and coordinate grid where
$J_{\nu}(z)$ is defined, and linearize the  Planck function, $B_{\nu}$,  with respect to
$T_{\rm d}$. Equation~\eqref{MainEq} is approximated by a set of finite difference
equations for each $z$-grid point, while Equation~\eqref{EEBFinal} is represented by a finite
sum. As a result, we get a system of linear equations for $J_{\nu_i}(z_k)$  that can be
written  using a hypermatrix  formalism. This hypermatrix system is solved with the tridiagonal
Thomas algorithm  \citep{Press:1992p7514}. After the new values of $J_{\nu_i}(z_k)$ and $T_{\rm d}(z_k)$ are
obtained we refine linearization for the  Planck function, update the system, and repeat
iterations until convergence  is achieved.

The stellar and diffuse interstellar radiation fields can be treated as boundary
conditions to the above system  of equations. We  use an approach developed by
\citet{Dullemond:2002} and consider stellar and interstellar fields as non-radiative
additional source terms in Equation~\eqref{EEBFinal}. This approach takes into
account the shielding of the star by  the inner parts of the disk.  For that one
needs to know the fraction of stellar radiation intercepted by the disk at each radius. We
compute the corresponding grazing angle as an angle between  dust
density isoline at $\rho_{\rm d}=5\cdot10^{-24}$ \gcmc and the direction toward the star. For the stellar spectrum, we use a
4000\,K blackbody for $\lambda>4000$\,\AA\,.  For shorter wavelengths, we use the
interstellar radiation field \citep{Draine1978,DB96} with an extension to longer wavelength
\citep{vanDishoeck1982}, where we have scaled the intensity so that it is continuous at
the transition wavelength of 4000\,\AA\,. Such a normalization leads to typical values of
stellar UV intensity at disk atmosphere being equal  to $\sim500$ ``Draine units``
\citep{Rollig2007} at a radius of 100\,AU. 

\subsubsection{Dust opacities and size distributions} 
 As a result of dust evolution  modeling we get dust size distribution functions $f(a,R,z) [\rm
cm^{-4}]$ being the fraction of grains with sizes within $(a, a+da)$ interval. To compute
dust opacities  one should know efficiency factors for dust absorption $Q_{\rm abs}$ and
scattering $Q_{\rm sca}$: 
\begin{equation} 
\kappa_{\nu}=\int\limits_{a_{min}}^{a_{max}}\pi a^2 Q_{\rm abs}(a,\nu) f(a)\,da. 
\end{equation} 
\begin{equation}
\sigma_{\nu}=\int\limits_{a_{min}}^{a_{max}}\pi a^2 Q_{\rm sca}(a,\nu) f(a)\,da.
\end{equation} 
$Q_{\rm abs}$ and $Q_{\rm sca}$ are computed from the Mie theory for
astrosilicate grains \citep{Laor1993}, but any other opacity model can be easily adapted.

%% file: GTB.tex
\newcommand{\htot }{\ensuremath{n_{\rm H}}}

%%%%%%%%%%%%%%%%%%%%%%%%%%%%%%%%%%%%%%%%%%%%%%%%%%% 
\subsection{Gas thermal balance}
\label{ThermBal} 
%%%%%%%%%%%%%%%%%%%%%%%%%%%%%%%%%%%%%%%%%%%%%%%%%%%%
%--------------------------------------------------------------- 
The kinetic gas temperature $T_{\rm g}$ is obtained by solving the thermal balance equation:
\begin{equation} 
\sum_k \Gamma_k(T_{\rm g}, T_{\rm d}, \rho_i) - \sum_k \Lambda_k(T_{\rm g}, T_{\rm d}, \rho_i, n_j^{\rm sp}) = 0\,, 
\end{equation} 
where $\Gamma$ and $\Lambda$ are  gas heating and cooling rates in $\rm erg\,s^{-1}\,cm^{-3}$. They
depend on  absolute abundances of main heating/cooling species $\rho_i$ and their level populations
$n_j^{\rm sp}$, which in turn depend on the gas temperature. Therefore, the problem is
solved iteratively at each grid point, starting from the disk atmosphere boundary toward
the midplane for any given radius, by means of the Brent method \citep{Press:1992p7514}.
%**************************************************************************

%--------------------------------------------------------------- 

Stellar FUV radiation is the main  gas heating source  in
protoplanetary disks, leading to a  PDR-like  structure of the upper disk
regions. There, gas is mainly heated via the photoelectric  (PE) effect on dust grains
and PAHs. In addition, collisional de-excitation of H$_2$ pumped by FUV photons,
photodissociation of H$_2$, and carbon  photoionization are important heating sources in
 specific disk regions. Gas heating by exothermic chemical reactions plays only a minor
role, with the largest contribution coming from H$_2$  recombination on
grains. In the optically thick, dense disk interiors, the dominant heating sources are
 the cosmic ray ionization of H and H$_2$, and  viscous heating due to dissipation of
accretion energy. Gas mainly cools via non-LTE atomic and molecular line emission, collisions
with grains, and, at high temperatures, by
emitting Ly$\alpha$ and metastable line emission.  The details of heating and cooling processes
can be found in Appendix~\ref{AppendA}.
%---------------------------

%--------------------------------------------------------- 

%--------------------------------------------------------------- 
\subsection{Chemistry}
\label{chemistry} 
An important ingredient of the thermal balance calculations is chemistry.
 While often a fast, simplified equilibrium approach is adopted, 
time-dependent chemical modeling may be more appropriate for calculations of abundances of major molecular
coolants. We adopted the same gas-grain chemical model as in \cite{Vasyunin2011}.
The reactions and reaction rates are based on the RATE'06 chemical
ratefile \citep{2007A&A...466.1197W}. For all photochemical
reaction rates, we use the local mean intensity (as a
function of $\nu$) computed with the RT model. To compute photoreaction rates, the
dissociation and ionization cross-sections from \citet{vDea_06} and the AMOP
database\footnote{\url{http://amop.space.swri.edu/}} are utilized. If cross-sections are not available
for a certain reaction, we retain the standard $\chi_0 \exp(-\gamma A_{\rm V})$ formalism,
with a $\gamma$ value taken from RATE'06 ratefile, $\chi_0$ estimated at the  upper disk
boundary, and $A_{\rm V}$ computed as $\ln(\chi/\chi_0)$. The same values are used to
estimate photodesorption  rates. Thus, the  calculation of photoprocesses takes into account the detailed shape of the
incident UV spectrum of the central star and its penetration through the disk.
Self-shielding for H$_2$ dissociation is computed using the \citet{DB96} formalism, with
the modified $A_{\rm V}$ values used to account for dust attenuation. The self- and mutual
shielding for CO photodissociation are computed using recent tabular data of
\citet{2009A&A...503..323V}.

The unattenuated cosmic ray (CR) ionization rate is assumed to be $1.3\times10^{-17}$\,s$^{-1}$. The
surface reactions are taken from \citet{Garrod_Herbst06} and assumed to proceed without
hydrogen tunneling via potential wells of the surface sites and reaction barriers. Thus
only thermal hopping is a source of surface species mobility. A diffusion-to-desorption energy ratio of 0.77 is adopted for all species  \citep{2000MNRAS.319..837R}. Under these
conditions, stochastic effects in grain surface chemistry are negligible, and classical
rate equations may be safely used \citep{2009ApJ...691.1459V,2009ApJ...700L..43G}. As the
initial abundances, we utilize a set of ``low metals'' neutral abundances from
\citet{Lea98}, where most of refractory elements are assumed to be locked in dust grains.

As the density and temperature distributions, computed here, are similar to those used in
\cite{Vasyunin2011}, we decided to use the same vertical distributions of X-ray ionization
rates regarding them as reference values. In the chemical module they are simply added up
to CR ionization rates. For the purpose of chemical evolution, we assume that dust is
represented by grains with a single size which is computed from the local grain size
distribution as described in \cite{Vasyunin2011}.
 With this chemical model, a disk chemical structure is computed
using dust properties and physical conditions from the previous iteration.
We assume that the grain properties do not change over the computational time span of 2~Myr.

%% file: HS.tex
\subsection{Vertical gas distribution}

Given  that the gas temperature $T_{\rm g}(R,z)$ and the mean molecular weight $\mu(R,z)$ are known, 
the vertical gas density distribution can be found by integrating the hydrostatic equilibrium equation:
\begin{equation}
 \frac{\partial P(R,z)}{\partial z}=-\rho(R,z)\frac{GM_{\star}z}{\left(R^2+z^2\right)^{3/2}}, \label{GS}
\end{equation}
coupled to the equation of state of  the ideal gas:
\begin{equation}
  P(R,z)=\frac{kT_{\rm g}(R,z)}{m_{\rm p}\mu(R,z)}\rho(R,z) \label{ES},
\end{equation}

In this study we assume that the radial profile of the surface density is given by the known function 
$\Sigma(R)$. 

%% file: DE.tex
\subsection{Dust evolution}

The evolution of the dust size distribution is calculated using the model presented in
\citet{Birnstiel:2010p9709}.  In this work, we consider neither the viscous evolution of the gas disk nor the radial
evolution of the dust surface density.

The grain evolution begins with grains sticking at low
collision velocities. Disruptive collisions at higher impact velocities cause erosion or
fragmentation, which poses an obstacle towards further  grain growth  and replenishes  the 
 population of small grains. Typical threshold collision velocities for the onset of fragmentation are
found to be about 1~m~s$^{-1}$ for silicate dust grains \citep{Blum2008p1920}. Icy
particles are expected to fragment at higher velocities due to the increased surface
binding energies \citep{Wada:2008p4903,Gundlach:2011p15761}. We therefore use a threshold
velocity for fragmenting collisions of 10 and 30~m~s$^{-1}$ in our dust models. Grain collisions are driven by
relative velocities due to Brownian motion, turbulent motion \citep{Ormel:2007p801}, radial and azimuthal drift
as well as vertical turbulent settling.

In order to make the calculation of the dust evolution feasible, we consider a radially
constant, vertically integrated dust-to-gas ratio and an azimuthally symmetric disk. The
vertical settling of dust is taken into account by using a vertically integrated kernel
\citep[see][]{Brauer:2008p215,Birnstiel:2010p9709}. The integration implicitly assumes
that the vertical distribution of each dust species follows a Gaussian distribution with a
 size-dependent scale height. This is a good approximation for the regions close to
the disk midplane where coagulation is most effective.  However, for modeling of the chemical evolution
the detailed vertical distribution of each dust species needs to be known. We
therefore use a vertical mixing/settling equilibrium distribution
\citep{Dullemond:2004p390}, taking a vertical structure of the previously calculated dust surface
densities.

%% file: results_and_discussions.tex
\section{Results}

\subsection{Disk physical structure for evolved and pristine dust cases} 
As an initial
approximation, we adopt a disk from \cite{Vasyunin2011} with mass $M_{\rm
disk}=0.055M_{\odot}$ and gas surface density profile $\Sigma(R)$ close to a power-law
with index $p=-0.85$ and $\Sigma(1\rm AU)= 34$~g/cm$^2$. The dust surface density is equal to
 $1\%$ of the gas surface density. We assume the following parameters  for a central star:
a mass $M_{\star}=0.7 M_{\odot}$, a radius $R_{\star}=2.64 R_{\odot}$ and an effective
temperature $T_{\star}=4000$~K.  This system closely resembles the DM~Tau disk. As UV-excess we use
the standard interstellar diffuse radiation field \citep{Draine1978} and its extension to longer wavelengths
\citep{vanDishoeck1982} as described in Section~\ref{MainEquationRT} (``JD'' case from
\cite{Akimkin2011}). To show the influence of dust evolution on the disk thermal and
density structure we present results for two cases:  the pristine dust with uniform
distribution and grain size of $0.1 \rm\mu m$ (Model A) and  the dust distribution and sizes as
obtained with the dust evolution model after 2~Myr of evolution  (Model Ev). The maximum grain size, attained in the midplane in Model Ev, varies from $4\cdot10^{-3}$~cm at 550~AU to 0.02~cm at 10~AU. The minimum grain size is always $3\cdot10^{-7}$~cm.

\begin{figure}[!ht]
\includegraphics[scale=0.3,angle=0]{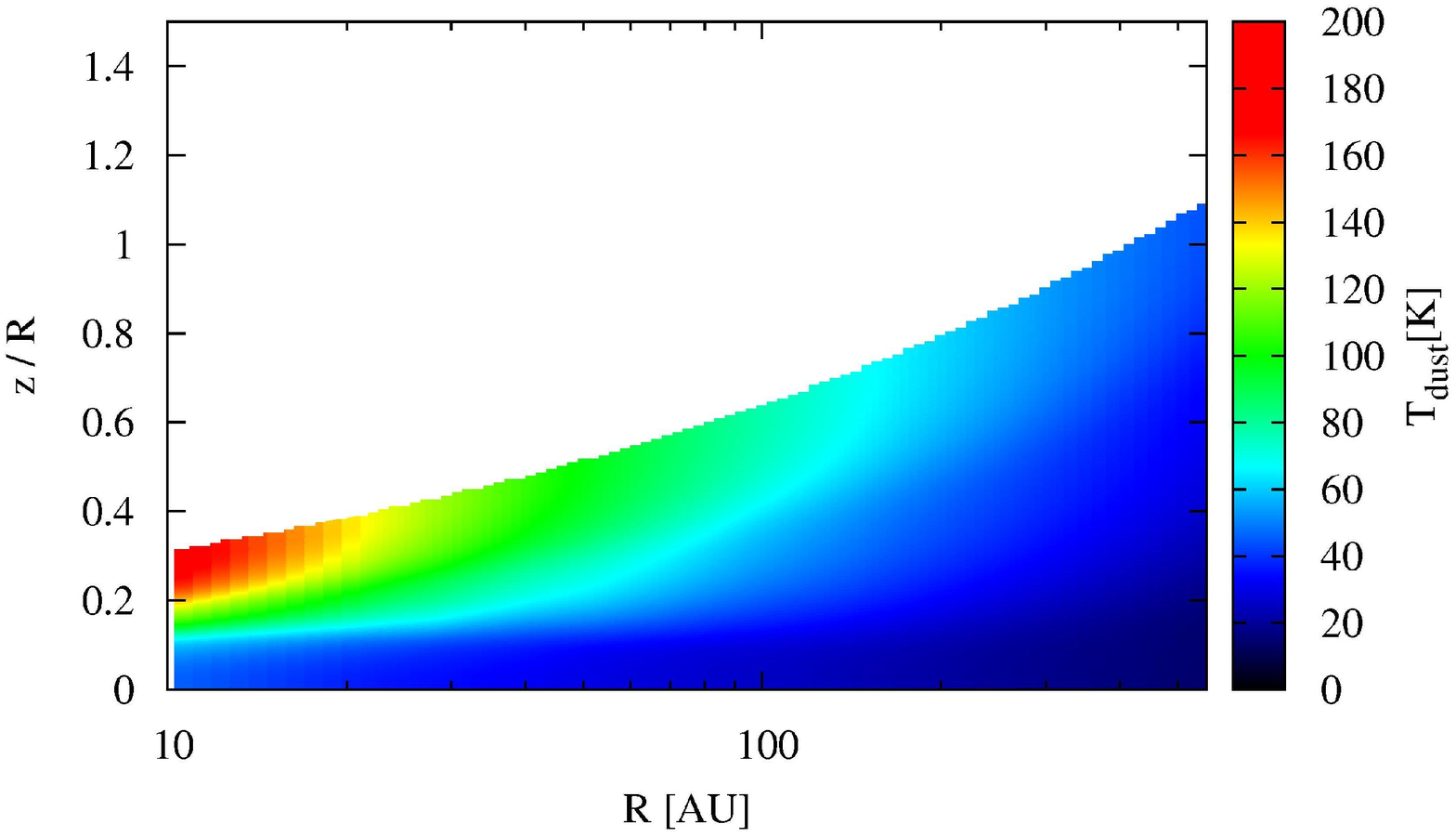}
\includegraphics[scale=0.3,angle=0]{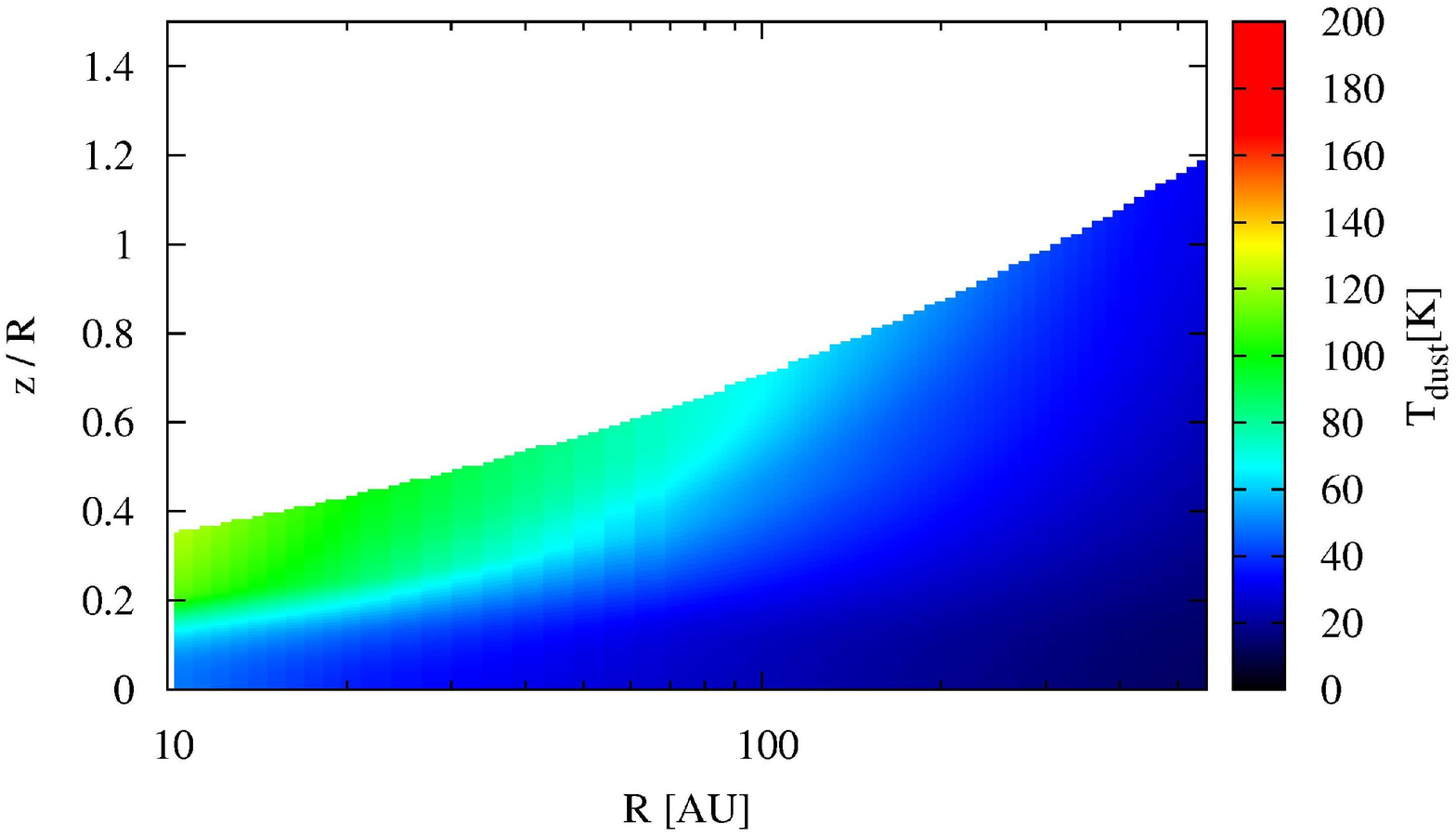}
\caption{ Dust thermal structure for  the disk model with the evolved (left panel) and pristine well-mixed (right
panel) dust.} 
\label{2D_Td} 
\end{figure}

In Figure~\ref{2D_Td} the dust temperature distribution is shown for  the both cases.  The disk model with
evolved dust is
hotter by about 70~K in the inner disk atmosphere ($R<200$~AU) and by $\sim 10-20$~K in the outer
atmosphere ($R>200$~AU)  compared to the disk model with the pristine dust,
whereas the dust midplane temperatures are quite similar in  the both
cases.  Higher dust temperatures at the disk surface in the evolved dust model are explained by a steeper
slope of  dust opacities in the mid-IR, where  such dust predominantly emits.
 Since both disk models have similar dust midplane temperatures and  due to transparency of the  bulk disk
matter to the far-IR/millimeter emission, the emergent disk spectral energy distributions (SED) are similar.
The difference in emergent spectra between Model~A and Model~Ev becomes apparent mostly at mid-IR wavelengths,
where dust continuum emission from the inner disk parts peaks.

\begin{figure}[!ht]
\includegraphics[scale=0.3,angle=0]{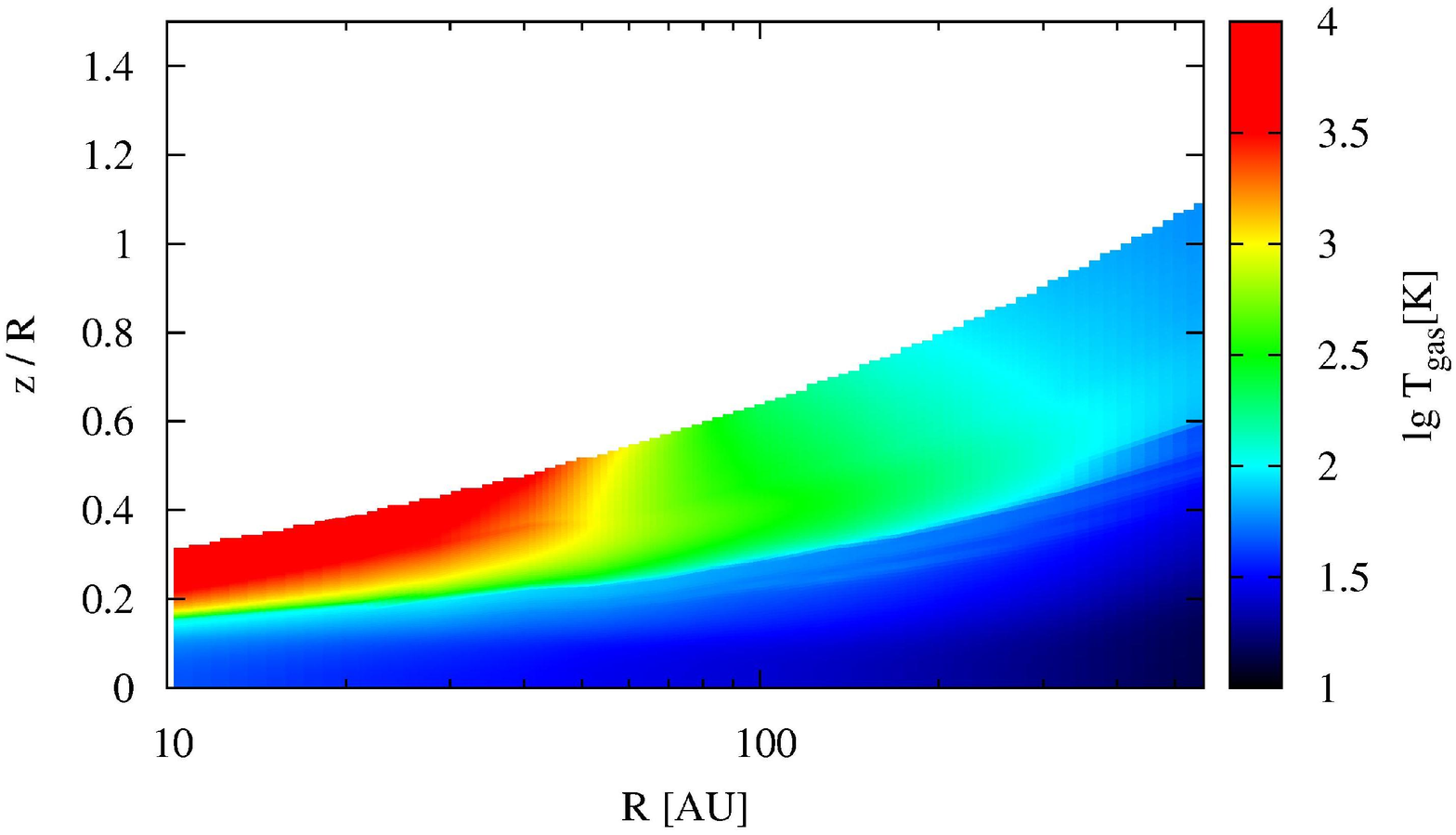}
\includegraphics[scale=0.3,angle=0]{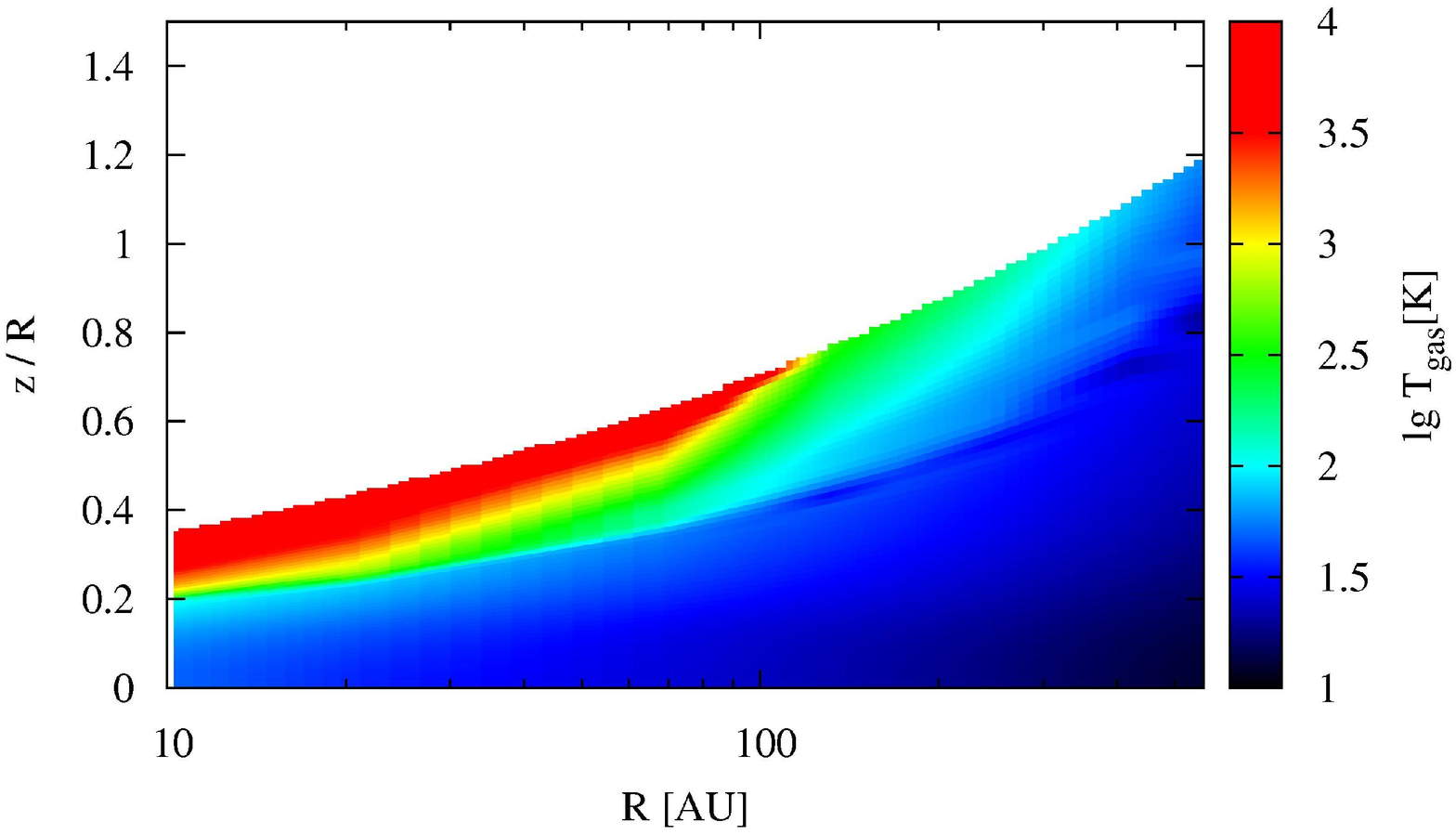} 
\caption{ Gas thermal structure for  the disk model with the evolved (left panel) and pristine well-mixed (right
panel) dust.} 
\label{2D_Tg} 
\end{figure}

 Gas temperature distributions in the disk models with the evolved and the pristine dust are shown in
Figure~\ref{2D_Tg}  and can be compared with the dust temperatures in Figure~\ref{2D_Td}.
The extent of the gas-dust thermal coupling zone  (where $T_{\rm d} = T_{\rm g}$)
in Model~Ev is slightly smaller  than in Model~A, primarily due to
sedimentation. As the midplane dust  temperatures for the two models do not significantly
differ, the gas temperatures also inherit this behavior.  On the other hand, the gas temperature
distributions above the coupling zone are quite different. In  the both cases, the inner disk
atmosphere is heated up to several thousand  Kelvin by photoelectric heating, but it is $\gtrsim
1000$~K cooler in Model~Ev. This is due to the reduced
abundance of grains in  the Model~Ev, where the main contribution to  the photoelectric heating
comes from PAHs.

\begin{figure*}[!ht] 
\centering
\includegraphics[scale=0.9,angle=0]{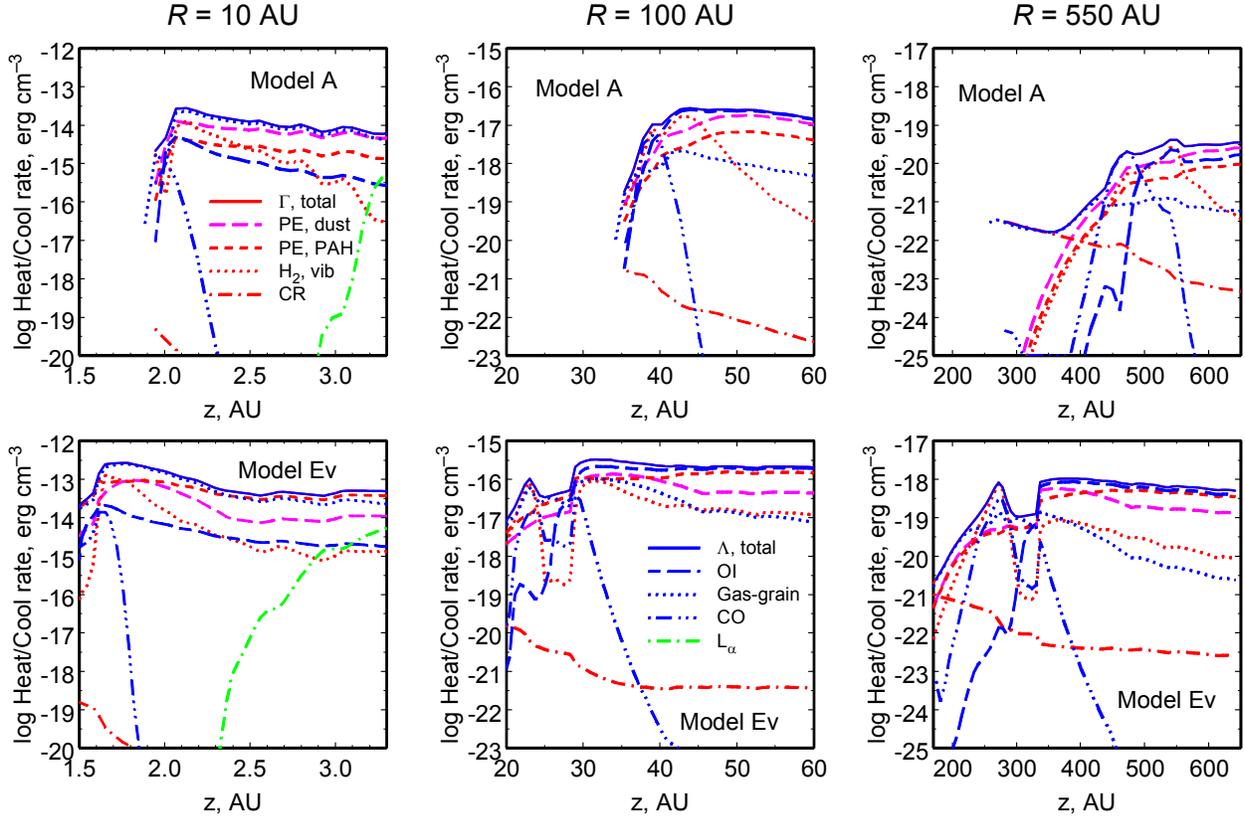}
\caption{Main gas heating and cooling processes at selected radii. To avoid confusion, the legend is split into two parts. Same line styles apply on all plots. To illustrate the thermal coupling zone extent the cooling and heating functions are not plotted in regions with $T_{\rm d}\cong T_{\rm g}$.} 
\label{hcatthreerad} 
\end{figure*}

 In contrast, in Model~A grains dominate photoelectric heating. Their intense heating
in the upper atmosphere has to be balanced by $Ly\alpha$ cooling, while in Model~Ev
remaining grains in the inner disk and  the [O~I] line cooling at  larger distances
($\gtrsim 40$~AU) can balance  the photoelectric heating from PAHs. Radial extent of hot
tenuous atmosphere is drastically different for  the two disk models: it exceeds 100~AU in Model~A,
whereas in Model~Ev gas is cooler than 1\,000~K even at $R=60$ AU. Absence of grains in the
disk atmosphere in Model~Ev leads to  an increase of the gas  temperatures by about factor of 2 at $z/R$
between $\approx
0.3$ and 0.6 at $R\approx 100-300$~AU. Somewhat smaller  increase of the gas
temperatures  in the upper layers is also present in more distant regions of the disk. Rates of main heating and cooling processes are shown
in Figure~\ref{hcatthreerad}.

\begin{figure}[!ht]
\begin{flushleft} 
$ \begin{array}{cc}
\includegraphics[scale=0.3,angle=0]{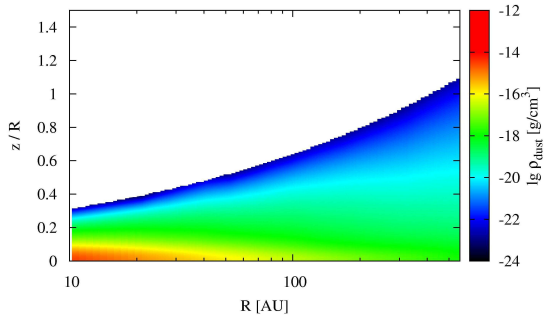} &
\includegraphics[scale=0.3,angle=0]{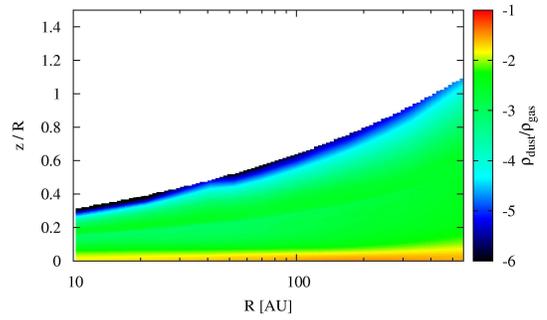} 
\end{array}$ 
\end{flushleft}
\caption{Distribution of  the dust density in Model Ev (left panel) and Model A (right
panel).} 
\label{2D_rhod} 
\end{figure}

\begin{figure}[!ht] 
\begin{center} 
\includegraphics[scale=0.3,angle=0]{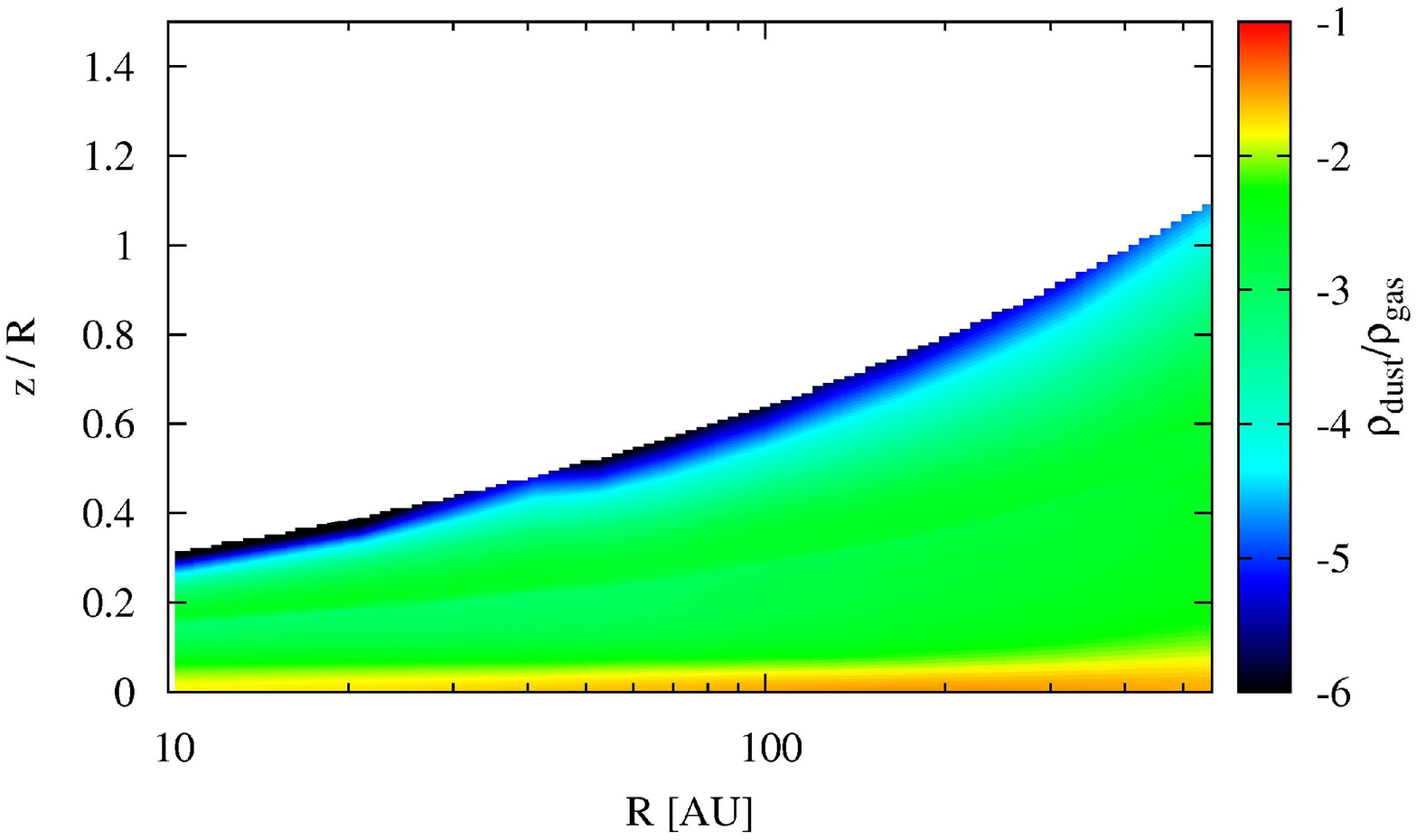}
\end{center} 
\caption{ Dust-to-gas ratio for Model Ev.}
\label{d2g} 
\end{figure}

\begin{figure}[!ht] 
\begin{center} 
\includegraphics[scale=0.3,angle=270]{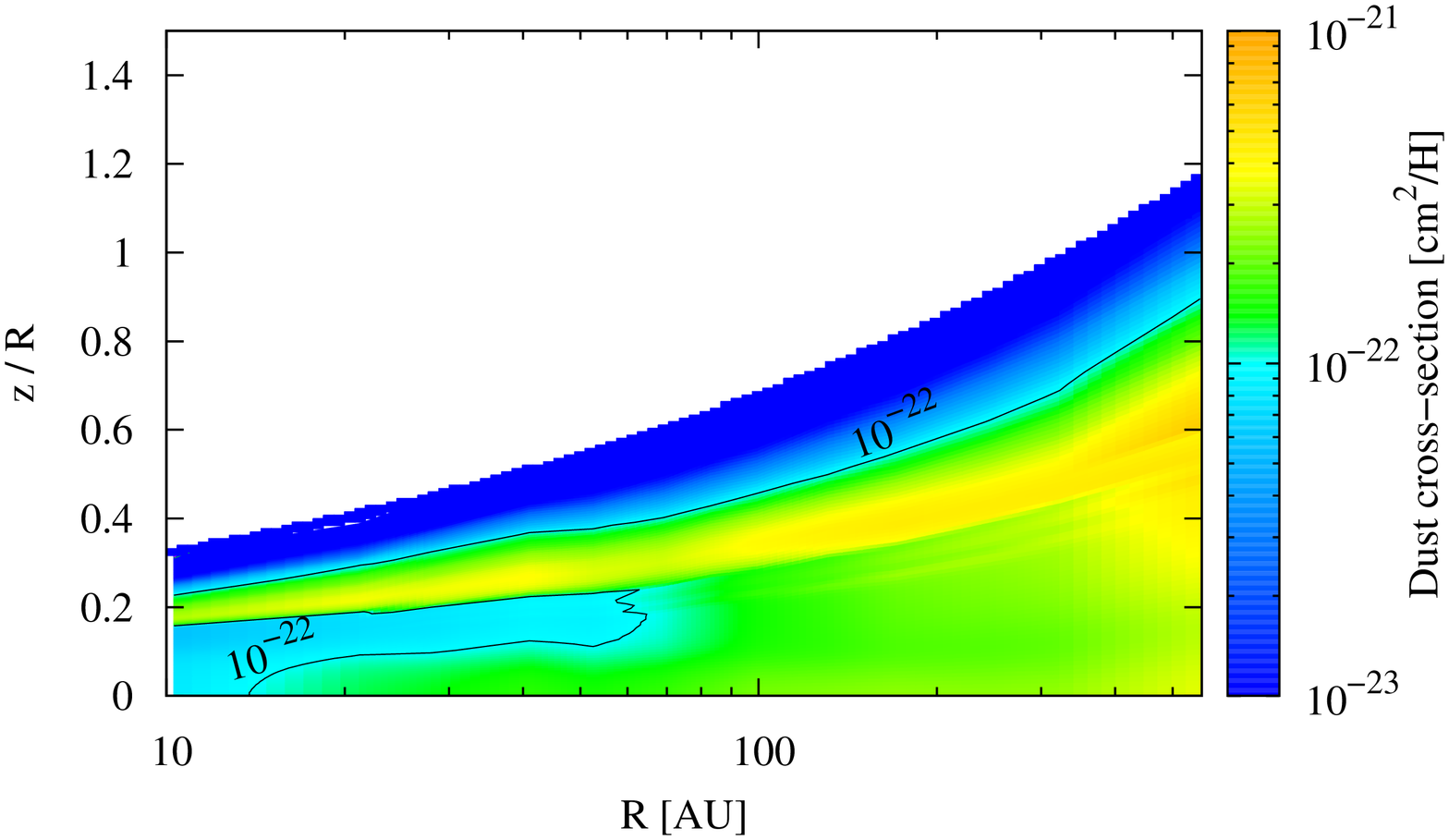}
\end{center} 
\caption{ Dust cross-section per hydrogen atom for Model Ev. The correspondent value for Model A is $5.9\cdot10^{-22}$ cm$^2$/H.}
\label{SperH} 
\end{figure}

Dust density distributions for Models~Ev and A are compared
in~Figure~\ref{2D_rhod}. The dust  densities differ by several orders of magnitude, mainly due to
sedimentation. This process also leads to dramatic changes in  the dust-to-gas ratio. While in Model~A this value
is constant and equal to $10^{-2}$, in Model~Ev we encounter the whole range
of values, from $10^{-1}$ to $ 10^{-8}$ (see~Figure~\ref{d2g}). However, dust-to-gas ratios
below $10^{-4}$ lead to unstable solutions  and poor convergence of the code,
therefore in the calculations we  assume that the minimal value of dust-to-gas ratio  is $10^{-4}$.
In Figure~\ref{SperH} the dust cross-section per hydrogen atom is presented for Model Ev. In case of Model A
this value is equal to $5.9\cdot10^{-22}$ cm$^2$/H everywhere in the disk.

\input chem.tex 
\input diss.tex

\input timechem.tex

%% file: chem.tex
\subsection{Chemical structure}

 One of the main goals of our study is to probe potential changes in the disk chemical
structure that may arise due to various processes related to the dust evolution. In the
section we present a detailed comparison of molecular abundances in  the disk models with pristine
and evolved dust for radii of 10~AU, 100~AU, and 550~AU. These are the same radii that
have been analyzed by \cite{Vasyunin2011}.  We consider only those
species that have mean abundances exceeding $10^{-12}$ at least in one of the two
considered models. The mean abundance is computed as a ratio of the species column density to
the column density of hydrogen nuclei. Remember that in all three cases we  ignore and do not show the
vertical structure at the height where the mass density of dust grains drops below the
adopted limit of $5\cdot10^{-24}$~g\,cm$^{-3}$ in the evolved model, and the medium can be
considered as purely gaseous. In the case of the well-mixed dust disk model, 
this value corresponds to the hydrogen number density of $2\cdot10^2$~cm$^{-3}$, 
and is even lower for the dust evolving disk model.

\begin{figure*} 
\centering
\includegraphics[width=0.7\textwidth,clip=]{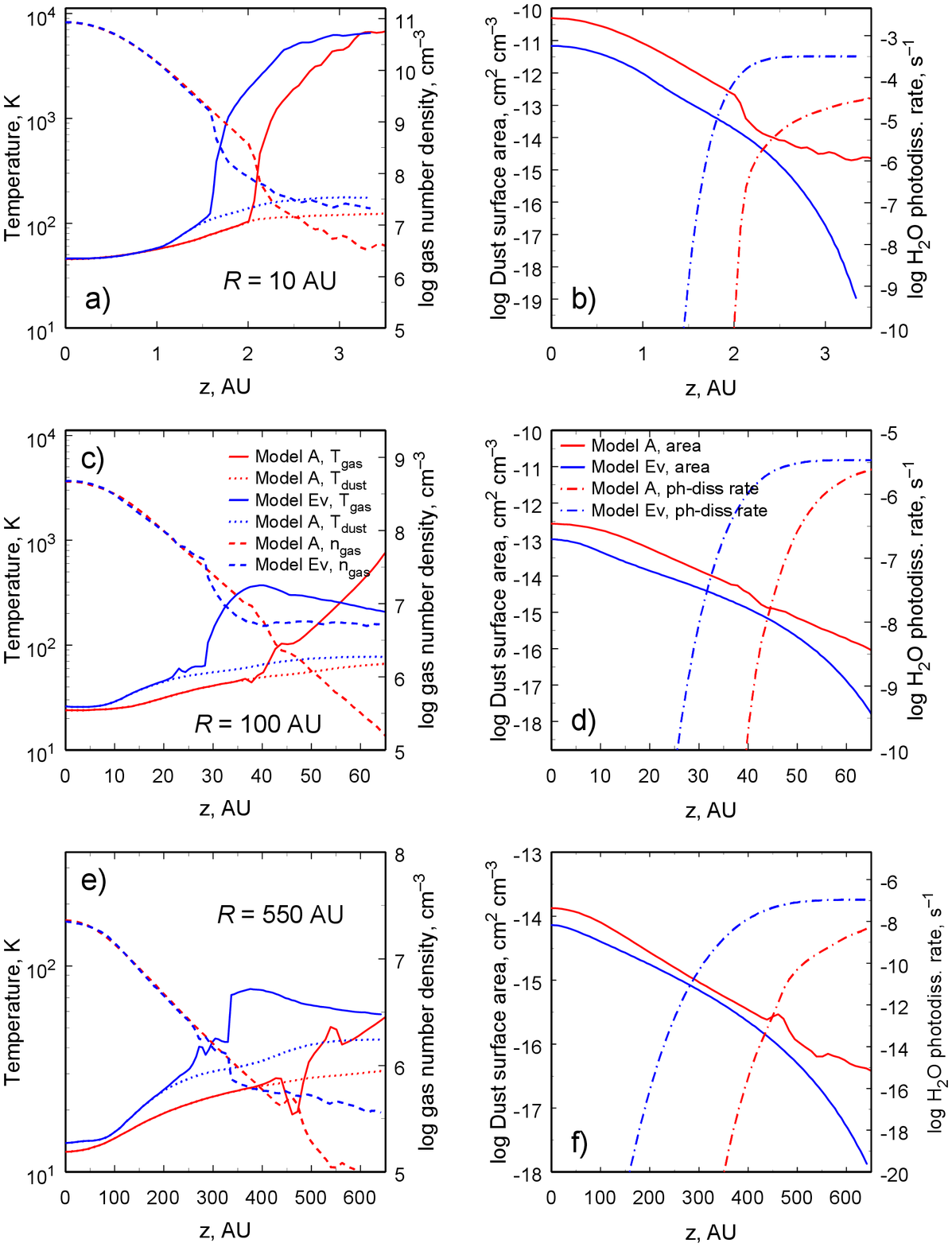}
\caption{Vertical distributions of selected disk parameters at three radii selected for a
chemical analysis as indicated in titles. In all plots red lines correspond to a model
with pristine dust, while blue lines correspond to a model with evolved dust. Shown in the
left column are gas densities (dashed lines), dust temperatures (dotted lines), and gas
temperatures (solid lines). Plots in the right column show dust surface area per unit
volume (solid lines) and water photodissociation rates (dot-dashed lines). }
\label{totstruct} 
\end{figure*}

The key disk properties at the selected radii are shown in Figure~\ref{totstruct}. Up to
a certain height the gas density is almost the same in both models. Above this height
the vertical density profile flattens in Model~Ev, because gas temperature either stops
increasing or decreases with $z$, and density stays nearly constant to keep the disk
hydrostatically stable.

 The main reason for that is the disk transparency.
The dash-dotted lines in Figure~\ref{totstruct} (b, d, f) show the water photodissociation
rates that are used here as a descriptive characteristics of the radiation field  strength.
Obviously, in Model~Ev  the UV radiation penetrates deeper into the disk. Dust is warmer
in this model than in Model~A almost everywhere in the disk.  In Model~Ev gas is also
significantly hotter that in the model with pristine dust in the more illuminated region that extends
approximately from 1.5~AU to 3~AU at
$R=10$~AU, from 30~AU to 50~AU at $R=100$~AU, and from 100~AU to 600~AU at $R=550$~AU.

We characterize the dust evolution using the total dust surface area per unit volume that
is shown in Figure~\ref{totstruct} (b, d, f)  (solid lines). It is smaller in the
evolved model  as both grain growth and sedimentation reduce the
total surface of dust grains. While in the midplane this
reduction is mostly caused by the growth of dust grains and is quite moderate, from an
order of magnitude at 10~AU to less than a factor of 2 at 550~AU, in the upper disk, where
sedimentation is important, the total dust surface area in Model~A is greater by orders of
magnitude than in Model~Ev. However, this difference may not necessarily be important for
chemistry as it is mostly confined to the illuminated disk regions where dust mantles are
evaporated anyway by  the UV photons.

\subsubsection{Outer disk}

We start the description of the disk chemical structure from the outer disk, where
only minor changes in the disk physical parameters are caused by the grain evolution.
Because of ineffective grain growth the total grain surface area per unit volume is nearly
the same in both models, except for the outermost disk atmosphere (Figure~\ref{totstruct}, f). 
Second, the dust temperature is quite low in the disk midplane, so surface
reactions with heavy reactants should be mostly suppressed there.

\begin{figure*} 
\centering
\includegraphics[width=0.8\textwidth,clip=]{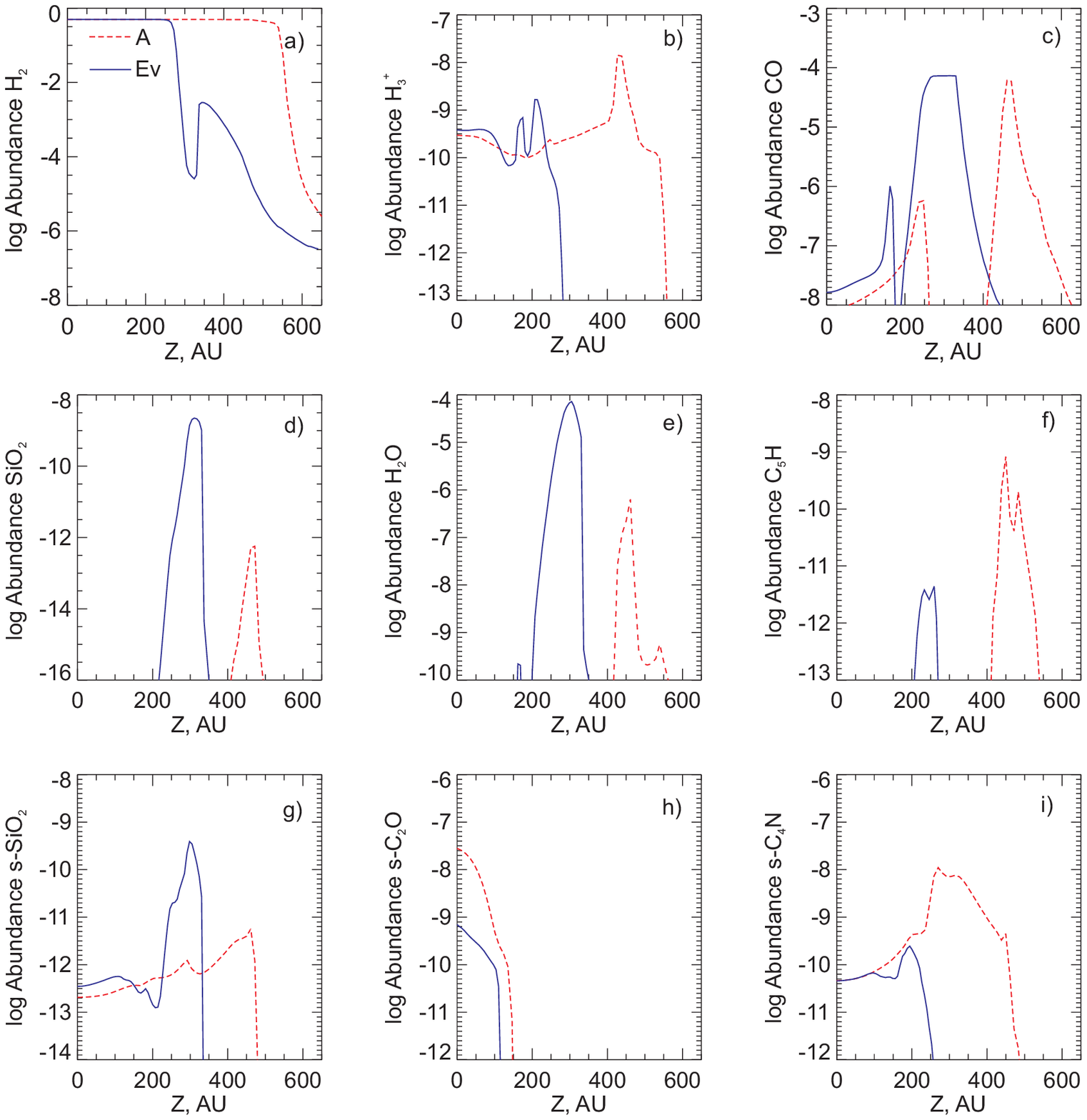}
\caption{Vertical abundance distributions of selected species at $R=550$~AU.}
\label{chem550} 
\end{figure*}

After 2~Myr of evolution we end up with 91 gas-phase species and 81 surface species that
have mean abundances greater than $10^{-12}$ either in Model~A or in Model~Ev. In most
cases grain evolution increases column densities for gas-phase components. Among the 91
gas-phase species only 13 have column densities that are smaller in Model~Ev than in Model~A.
The reason is quite straightforward.
As grains grow and settle down toward the
midplane, the so-called warm molecular layer moves down as well, to a denser disk region.
Even if relative abundances do not change significantly in the process, column densities
grow due to higher volume densities. Vertical abundance distributions for some species are
shown in Figure~\ref{chem550}.

 In the upper row of Figure~\ref{chem550} we present vertical abundance profiles for H$_2$, H$_3^+$, and CO. The key difference
between Model~A and Model~Ev is that in the former the warm molecular layer is located below the H$_2$ dissociation boundary,
while in the latter a portion of the molecular layer is located above this boundary, where free H atoms are abundant. This mutual
disposition is not impossible as the molecular layer and the H$_2$ photodissociation front are not directly related to each other.
Grain absorption of the FUV photons responsible for the H$_2$ dissociation is less significant in the model with evolved dust,
and the H$_2$ dissociation front is located deeper in Model~Ev. However, the overall transparency of evolved dust in the entire UV
range is smaller than for FUV photons only. Because of that the molecular layer that is controlled by desorption from dust
grains and dissociation of trace molecules is located somewhat higher. This specific result, of course, depends
on the adopted description of dust opacities and photoreaction rates.

Simple  atomic and diatomic components dominate the list of  the gas-phase species, whose
column densities are enhanced in Model~Ev. The largest column density increase at 550~AU
is found for SiO$_2$ (Figure~\ref{chem550}, d), N$_2$O, and water (Figure~\ref{chem550}, e).
In all cases it is related to higher abundance of a molecule in the molecular layer. Note
that almost all physical characteristics of the medium are nearly the same in the
molecular layers of the two  disk models, except for  the gas density and  the X-ray ionization rate. Higher
density in the molecular layer of Model~Ev accelerates two-body processes and shifts equilibrium
abundances of many molecules to higher values.

 The relative location of the molecular layer and the H-H$_2$ boundary, mentioned above, also plays a role in
defining molecular column densities, especially, for species that are produced in reactions with atomic hydrogen, like water. In
Model~Ev a significant portion of the molecular layer is located
above this transition, where abundant H atoms are available. This speeds up the gas-phase water synthesis in H~+~OH reaction as well as
surface synthesis of water molecules that are immediately released into the gas-phase due to photodesorption. This explains a huge
water spike located at height of about 300~AU. In Model~A water is mostly produced in surface reactions that are less effective
because of low H gas-phase abundance and lower rate of the UV photodesorption. Also, water
molecules are more rapidly destroyed in reactions with ions such as HCO$^+$ that are abundant in the
molecular layer due to higher X-ray ionization rate. The sharp drop-off in water abundance in both models coincides with the
carbon ionization front. Above the front, the main destruction routes for water molecules are the reaction with C$^+$
and photodissociation.

The situation is somewhat different for complex hydrocarbons, in particular, for long
carbon chains. Their abundances in Model~A are significantly enhanced in the molecular
layer in comparison with Model~Ev. This is again related to a more elevated position of
the molecular layer in Model~A. Because of that, it is less protected not only from  the UV
irradiation but from X-rays as well. Accordingly, ionized helium is more abundant in the
molecular layer of Model~A than in the molecular layer of Model~Ev. Abundant C-bearing
molecules, like CO, are destroyed by He$^+$ more efficiently in Model~A in the disk upper
region. Then, C$^+$ is consumed to produce simple CH$_{\rm n}^+$ species that stick to
grains and produce long carbon chains by surface processes. The dust temperature of the order
of 30~K is high enough to drive desorption of these molecules into the gas-phase.

This effect should not be overestimated. Even though the total column densities of carbon
chains are greater in Model~A than in Model~Ev, their absolute values are low, with the
mean abundance exceeding $10^{-12}$ only for C$_2$H, C$_4$, C$_4$H, C$_5$, C$_5$H, and
C$_6$H. The effect is most pronounced for C$_5$H (Figure~\ref{chem550}, f), with the ratio
of column densities in Model~A and Model~Ev of 15. For the observed C$_2$H
molecule the higher relative abundance in Model~A (related to more effective He$^+$ chemistry)
is nearly compensated by the higher absolute abundance in Model~Ev (related to deeper location
of the molecular layer), so its column densities are nearly the same in both models.

The behavior of surface species is different in the two disk models. While column densities of gas-phase species are
increased by grain evolution, column densities of many surface species decrease. There
are 81 abundant surface species at 550~AU, and only 30 of them have greater column
densities in Model~Ev. Also the difference of column densities of surface species is quite
modest in the two models.  Only for ten of 81 species column densities differ by more than
a factor of 3. Dominant surface carbon compounds (in terms of column densities) in both
models are carbon monoxide and methane. Because of low dust temperature, s-CO$_2$
production is suppressed, and this molecule in neither model reaches the high abundance
seen at smaller radii (see below).

Surface species that have greater column densities in Model~Ev are mostly silicon and
phosphorus compounds, which are not involved in surface chemistry 
(relevant reactions are not included in our chemical network). Their
abundances are
enhanced in the `main' molecular layer as are abundances of their gas-phase counterparts
(Figure~\ref{chem550}, g).

 Abundances of some surface carbon chains are enhanced in Model~A by about an order of
magnitude due to more intense X-ray ionization than in Model~Ev (see above). Also,
species like s-C$_2$O (Figure~\ref{chem550}, h) and s-C$_2$N involved in surface chemistry
have greater column densities in Model~A because their
midplane abundances are higher in this model due to greater available surface area  for their synthesis.
Carbon chains not involved in surface chemistry  in our chemical model, like s-C$_4$N (Figure~\ref{chem550}, i), mirror
 evolution of their gas-phase counterparts and have higher abundances in the upper carbon chain layer
described above.

\subsubsection{Intermediate disk}

As we move closer to the star,  at distances of about 50--100~AU, the fingerprint of dust evolution becomes more pronounced.
While the mass density of dust is greater in the midplane of Model~Ev due to
sedimentation, the total surface area is still 2.5 times less than in Model~A. In the
upper disk the difference in the surface area reaches a factor of 70. It is interesting to
note that the uppermost disk atmosphere is actually colder in Model~Ev than in Model~A,
despite being more transparent (Figure~\ref{totstruct}, c). This is because dust is not only
the main source of opacity but also an important heating agent (due to photo-effect). As
dust is depleted in the upper disk, the equilibrium temperature shifts to lower values,
dictated by  the PAH heating.

\begin{figure*} 
\centering
\includegraphics[width=0.8\textwidth,clip=]{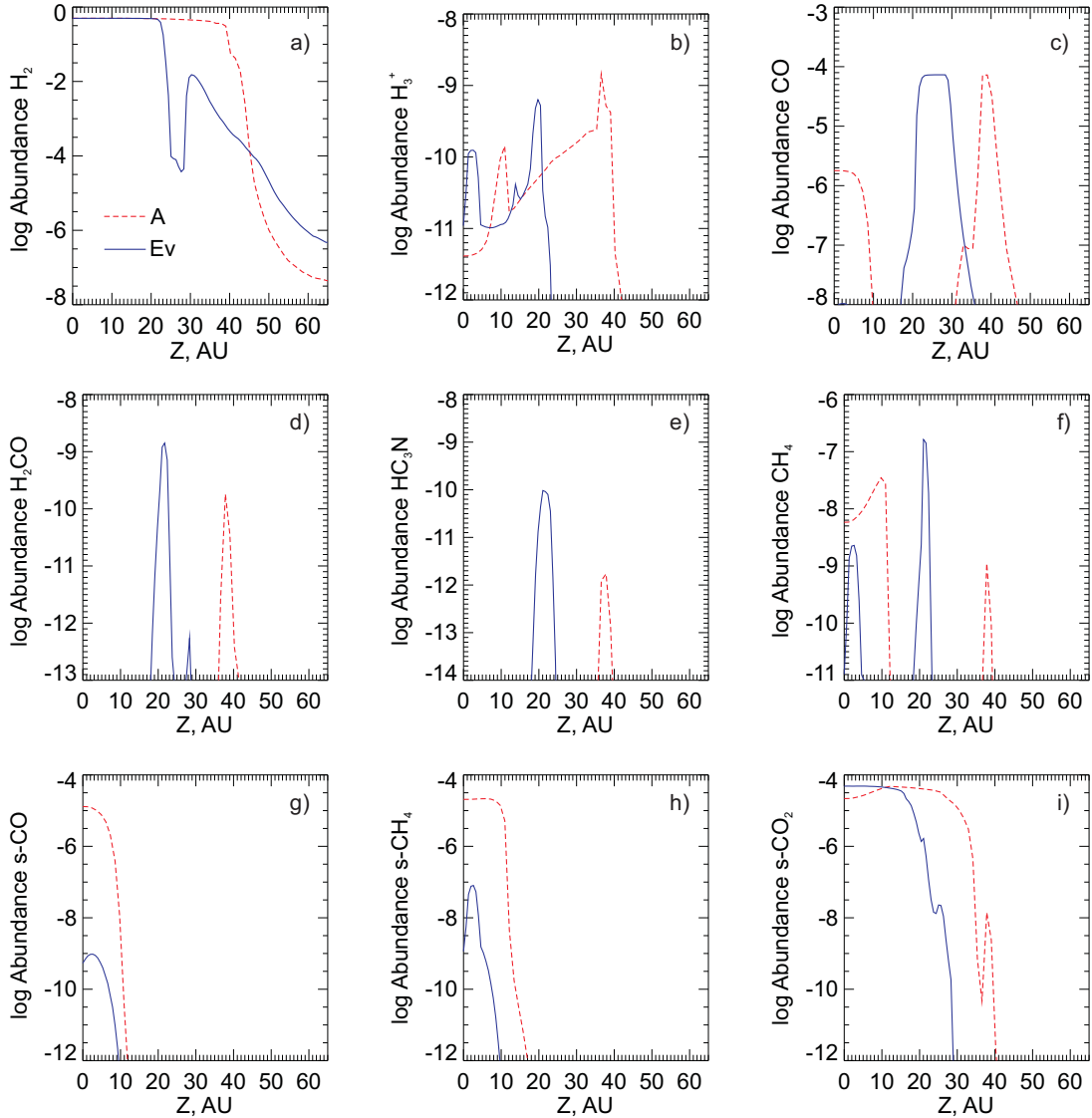}
\caption{Vertical abundance distributions of selected species at $R=100$~AU.}
\label{chem100} 
\end{figure*}

At $R=100$~AU, among 78 gas-phase species, having mean abundances higher than $10^{-12}$ at least in one
of the two models, most species (72) have higher column densities in Model~Ev, as at
$R=550$~AU, but the list of these species is somewhat different. Some examples of vertical
abundance profiles for gas-phase species at $R=100$~AU are shown in the top and middle row of Figure~\ref{chem100}.

 The main features of the chemical structure are the same as at 550~AU. In Model~Ev the molecular layer,
as marked by the CO distribution (Figure~\ref{chem100}, c), is located above the H$_2$
photodissociation front (Figure~\ref{chem100}, a). In Model~A the situation is the opposite. Also, in Model~A ions,
like H$_3^+$ (Figure~\ref{chem100}, b) are somewhat more abundant in the molecular layer which further
decreases abundances of neutral unsaturated molecules.

The largest difference between the two models is again observed for SiO$_2$ that has a
column density $3.6\cdot10^8$~cm$^{-2}$ in Model~A and $1.8\cdot10^{12}$~cm$^{-2}$ in
Model~Ev. Grain evolution causes water column density to increase from
$7.1\cdot10^{13}$~cm$^{-2}$ to $5.2\cdot10^{16}$~cm$^{-2}$. This is again related to the
different arrangement of the molecular layer and H--H$_2$ transition in Model~A and Model
Ev (Figure~\ref{chem100}, a).  The upper boundary of the water layer is defined by the location of the C ionization front.

Many complex gas-phase hydrocarbons, like formaldehyde (Figure~\ref{chem100}, d) and
cyanoacetylene (Figure~\ref{chem100}, e), are also affected. Among more or less abundant
molecules the only exception to this rule is methane (Figure~\ref{chem100}, f), with column
density being about 3 times larger in Model~A than in Model~Ev. This difference is related
to the surface chemistry as we will explain below.

The  chemical evolution of surface species is  complicated as  it is affected by at least two
competing factors related to grain growth. Less grain surface is available to the mantle
formation in Model~Ev, but because of higher grain temperature surface species are more
mobile than in Model~A, which intensifies the surface  recombination. The interplay between
these processes causes various responses of surface species to the grain evolution.

Thirty five of 72 abundant mantle components have higher column densities in Model~Ev.
Only for 24 species the ratio of column densities in the two models exceeds a factor of 3.
Higher abundances in Model~Ev are mostly typical for complex surface molecules with large
desorption energies that  have accreted from the gas phase and are not involved in surface chemistry.
These species mirror the
behavior of their gas-phase counterparts. Two striking examples of greater column
densities in Model~A are presented by s-CO and s-CH$_4$ (Figure~\ref{chem100}, g and h). Surface methane is underabundant by about 3 orders of magnitude in Model~Ev, while
surface CO is underabundant by more than 4 orders of magnitude in this model. As surface
species are mostly concentrated in the midplane, to explain this difference we need to
consider chemistry in this disk region.

Surface methane chemistry is quite limited in the adopted network. Methane is synthesized in a
sequence of hydrogen addition reactions, while the only effective route of its removal
from the mantle is desorption. So, its abundance is regulated by the balance between
hydrogen addition and desorption. As the desorption energy of methane is not very large
(1300\,K), desorption wins in this competition, and the steady-state s-CH$_4$ abundance in
Model~Ev shifts toward lower values. This does not work for other saturated molecules
(like water and ammonia), as they have much higher desorption energies, so their midplane
abundances in both models are just (nearly) equal to  the abundances of  the corresponding atoms.

Carbon monoxide is different from methane in the sense that it is not a `dead end' of some
surface chemistry route, but rather an intermediary on the route to s-CO$_2$ formation and
synthesis of complex organic molecules like CH$_3$OH. The reaction between s-OH and s-CO,
leading to s-CO$_2$, has a 80\,K barrier. This implies very strong dependence on the dust
temperature at critical $T_{\rm d} < 20-40$~K. Because of this dependence,
s-CO$_2$ formation is much more efficient in Model~Ev. While s-CO$_2$ is the most abundant
carbon compound in the disk midplane in both models, in Model~A its abundance is about two
times lower than in Model~Ev (Figure~\ref{chem100}, i). Thus, in Model~A, carbon atoms that
are not bound in s-CO$_2$ are available for other surface processes and are almost equally
distributed between surface methane, surface CO, and some other species, with s-CH$_4$
having almost the same abundance as s-CO$_2$. In Model~Ev, s-CO$_2$ synthesis proceeds
faster, and this component becomes not only the dominant C carrier, but also the dominant
oxygen carrier, leaving almost no place for either surface or gas-phase CO.

 The described trends are mostly preserved at $R=50$~AU. At this radius, dust evolution also causes column densities of
abundant gas-phase species to increase (mostly due to enhanced photodesorption). Water, carbon dioxide, formaldehyde, and
cyanoacetylene are among species mostly affected. Gas-phase methane column density is nearly the same in both models.
Surface methane and CO ice are underabundant in Model~Ev at $R=50$~AU, but to a less degree than at $R=100$~AU. Dust temperatures
in Model~A and in Model~Ev are nearly equal at this radius, so that in both cases surface s-CO$_2$ synthesis is very effective,
decreasing s-CO and s-CH$_4$ abundances and leveling differences between the two models. Abundances at 50~AU will be further
discussed in the last subsection of Section~3.

\subsubsection{Inner disk}
\begin{figure*} 
\centering
\includegraphics[width=0.8\textwidth,clip=]{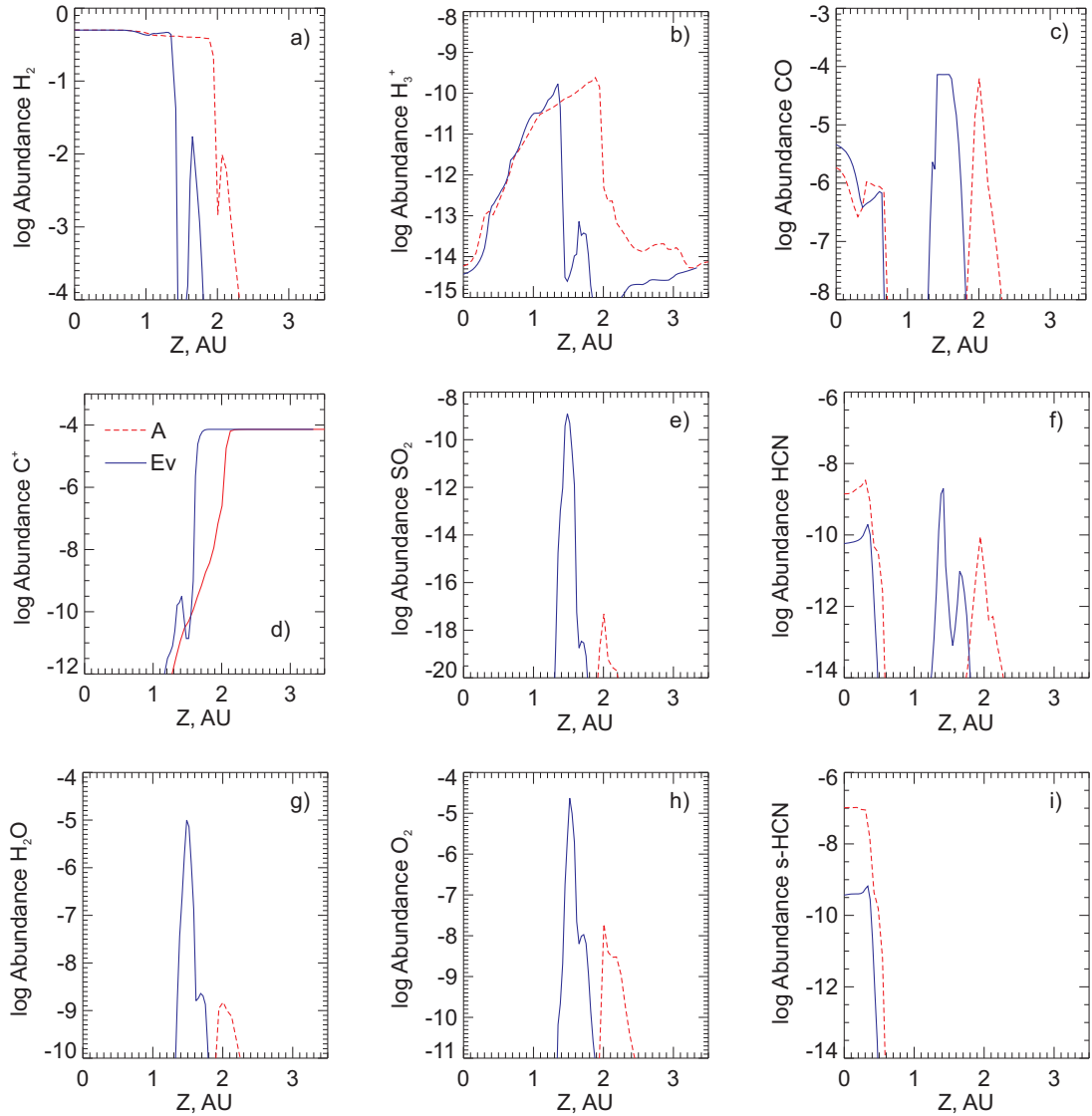}
\caption{Vertical abundance distributions of selected species at $R=10$~AU.}
\label{chem010} 
\end{figure*}

At $R=10$~AU we have 75 abundant gas-phase molecules, and 62 of them share the
common trend to be more abundant in the model with evolved dust. However, the magnitude of
the difference in column densities as well as its origin are related to other factors.
The molecular layers both in Model~A and in Model~Ev are located above the H-H$_2$
transition (Figure~\ref{chem010}, a). In both cases abundant H atoms are available both for surface and
gas-phase reactions. Despite the fact, water column density in Model~Ev exceeds that in
Model~A by more than 4 orders of magnitude (Figure~\ref{chem010},~g). This difference is
much greater than at other radii where we related it to the difference in atomic hydrogen
abundance. At these warm temperatures of $50-200$~K (see Figure~\ref{totstruct}, a), the formation of water is
dominated by neutral-neutral reaction
of O with H$_2$ (with the barrier of $1\,660$~K), followed by the neutral-neutral reaction of OH
with H$_2$ (with the barrier of $3\,163$~K), see \citet{2007A&A...466.1197W} and \citet{2011ApJ...743..147N}.

As H abundances in the molecular layers are nearly the same in both models, 
we need to find another explanation for the raise in water abundance in Model~Ev. It
is obviously related to the difference in physical parameters in the two molecular layers. Again, the
molecular layer in Model~Ev is shifted toward the midplane and, thus, resides in a denser
disk region. Because of higher density in the
molecular layer of Model~Ev, surface water synthesis is more effective there, increasing
its gas-phase abundance as well. Higher X-ray ionization rate in the molecular layer of
Model~A leads to higher ion abundances. In particular, it greatly enhances abundance
of H$^+$ that is one of the major water destroyers.  Another difference is the UV radiation spectrum that favors carbon
ionization in Model~A. In the adopted photoionization model, carbon atoms are ionized by the  UV radiation with wavelengths shorter
than 1100\,\AA. This radiation is absorbed less efficiently in Model A, and because of
that the C/C$^+$ transition zone is further vertically expanded, so that
the water layer falls in the region where C$^+$ abundance is still
significant (Figure~\ref{chem010},~d). This also leads to rapid water destruction.

Different water abundances cause even greater differences in column densities of SO and
SO$_2$. In the case of SO$_2$ the difference exceeds 9 orders of magnitude (Figure~\ref{chem010}, e). 
Significant growth of SO and SO$_2$ abundances can be traced to the
greater abundance of O$_2$ in Model~Ev. An SO$_2$ molecule is produced from SO, SO is
produced in reaction S + OH, atomic sulfur is the product of SO$^+$ dissociative
recombination, and SO$^+$ is produced in the reaction between S$^+$ and O$_2$. Abundance
of molecular oxygen in Model~Ev is greater by almost 4 orders of magnitude than in Model~A
(Figure~\ref{chem010}, h), which is also related to different H$^+$ abundances, as the H$^+$ +
O$_2$ reaction is one of the major O$_2$ destruction pathways. Thus, we conclude that at
$R=10$~AU chemical differences between Model~Ev and Model~A arise because grain evolution
shifts the molecular layer in the region of the disk that is more protected from X-rays and  FUV radiation.

Among species, that have their column densities decreased by grain growth, are HCN (Figure~\ref{chem010}, f) 
and HNC. They are mostly concentrated in the midplane, and their  midplane
abundances in Model~A exceed those in Model~Ev by an order of magnitude. Analysis of
chemical processes indicates that this difference is related to surface processes, that
is, higher gas-phase HCN abundance in Model~A simply reflects more effective surface
synthesis of the molecule because the available surface area is greater in this model (Figure~\ref{chem010}, i). 
Then, HCN desorbs into the gas-phase and gets  protonated by reactions with HCO$^+$ or H$_3^+$.
Dissociative recombination of HCNH$^+$ produces either HCN or HNC,
so the overabundance of HCN in Model~A is partially transferred into the overabundance of
HNC.

As for surface species, at this radius there are 64 abundant surface components, mainly heavy molecules. Nearly
half of them are more abundant in Model~Ev, but the increase in column densities is not
significant for most molecules. Two extreme examples of greater column densities in Model~A
are represented by s-HCN and s-HNC, for the reasons described above.

\subsection{Model with more efficient dust growth}
To check the sensitivity of our results to some details of the adopted grain physics, we considered additional models for dust evolution.
In this subsection we present a detailed description of Model Evx with a threshold velocity for fragmenting collisions increased from 10 to 30~m\,s$^{-1}$, which leads to more significant grain growth in the dense regions.

\begin{figure*}[!ht]
\centering
\includegraphics[width=0.6\textwidth]{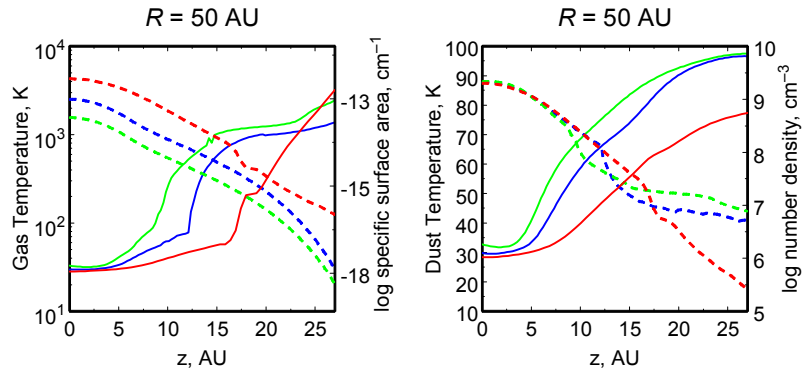}
\caption{Main disk structural properties at $R=50$~AU. In all cases red lines correspond to Model~A, blue lines correspond to
Model~Ev, and green lines correspond to Model~Evx. In the left panel gas temperature (solid lines) and dust surface area per unit
volume (dashed lines) are shown. Plotted in the right panel are dust temperature (solid lines) and gas number density (dashed
lines).}
\label{dev50struct}
\end{figure*}

Models~A, Ev, and Evx can be viewed as successive stages of the grain evolution process. So, in Model~Evx we may expect to see a
continuation of the same trends as were noted above for Model~Ev. In Figure~\ref{dev50struct} we show the main disk structural
properties at $R=50$~AU in the three models. Obviously, more advanced grain evolution causes the disk to become more transparent.
As a result, hot atmosphere becomes more extended, and dust becomes warmer, with the midplane grain temperature raising from 28~K
in Model~A to 33~K in Model~Evx. As we will see below, this relatively small difference has a noticeable effect on the disk
chemical structure. Dust surface, available for chemical reactions, is an order of magnitude smaller in Model~Evx than in Model~A.

\begin{figure*}[!ht]
\centering
\includegraphics[width=0.7\textwidth]{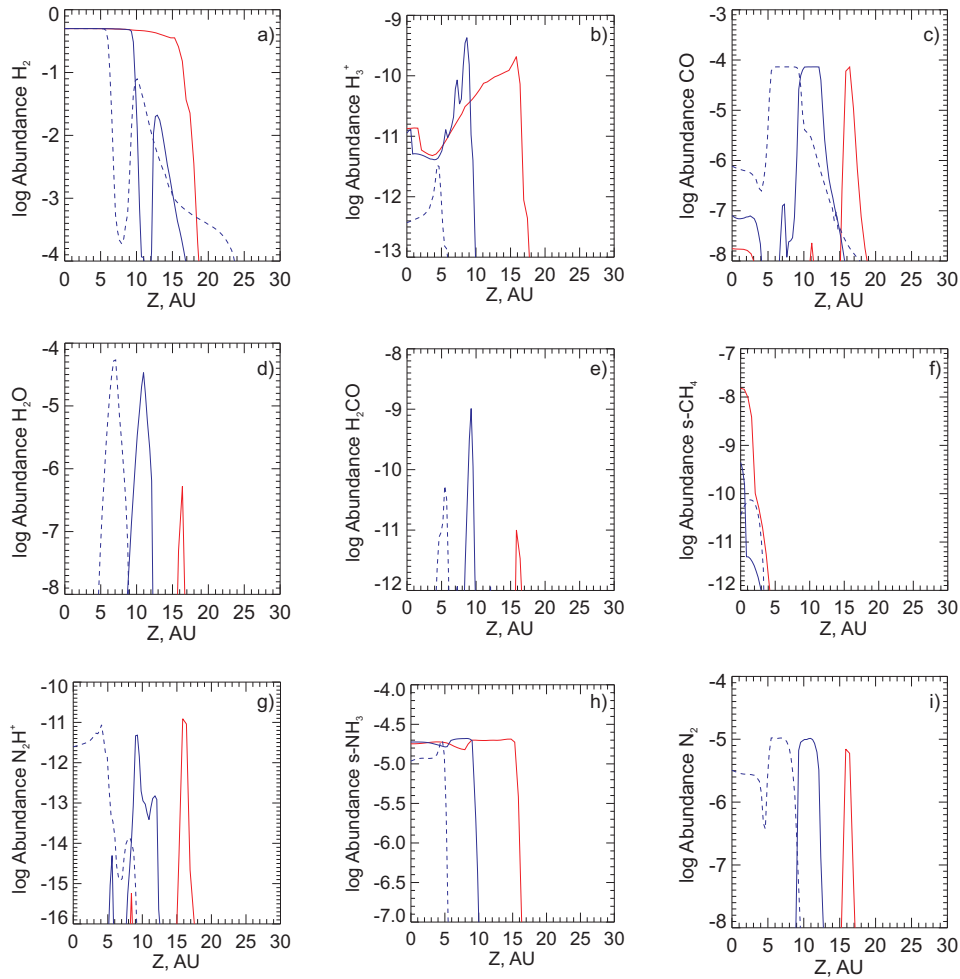}
\caption{Vertical abundance distributions of selected species at $R=50$~AU. Blue dashed line corresponds to the Model~Evx.}
\label{chem50}
\end{figure*}

Vertical profiles of some species for Models A, Ev, and Evx are shown in Figure~\ref{chem50}. As in previous cases, we start from
H$_2$ (Figure~\ref{chem50}, a) and notice that the H$_2$ photodissociation front sinks even deeper, so that hydrogen is almost
fully atomic above $\sim6$~AU. Due to warmer dust, gas-phase abundances of some molecules with low desorption energy are increased
in the midplane of Model~Evx (like in Model~Ev at $R=10$~AU). One can see the progressive growth of CO midplane abundance from
Model A to Model~Evx in Figure~\ref{chem50}~(c). Similar to CO, gaseous N$_2$ appears in the disk midplane in Model~Evx.
Protonation of such abundant  molecules lowers the H$_3^+$ abundance in the Evx model midplane (Figure~\ref{chem50}, b),
which affects abundances of some other ions, like H$_3$O$^+$.

A typical example of the molecular abundance evolution is shown in Figure~\ref{chem50}~(d). A peak of water abundance shifts toward
the midplane and grows higher. Due to increasing overall gas density and more intense photodesorption, gas-phase water
column density increases up to $4.7\cdot10^{17}$~cm$^{-2}$ in Model~Evx. The upper boundary of water layer is defined by the C$^+$
ionization front (Figure~\ref{chem50}, e).

A significant growth is observed for N$_2$H$^+$ column density. It increases from $4.8\cdot10^9$~cm$^{-2}$  to
$1.5\cdot10^{10}$~cm$^{-2}$ in Model~Ev and $5.1\cdot10^{11}$~cm$^{-2}$ in Model~Evx (vertical abundance profile is shown in
Figure~\ref{chem50}~(g)). This is related to increased thermal desorption of the N$_2$ ice and lower  abundances of
surface species that are mostly synthesized on grains, like methane (Figure~\ref{chem50},~f) or ammonia (Figure~\ref{chem50},~h).
In the latter case, some nitrogen atoms in the midplane are free to be incorporated into N$_2$ molecules (Figure~\ref{chem50},~i)
and further into N$_2$H$^+$ molecules. Species, significantly affected by the advanced grain growth, also include other gas-phase
molecules, related to surface chemistry. Column densities are increased by more than an order of magnitude in Model~Evx relative
to Model~Ev for H$_2$O$_2$, CH$_4$, CO$_2$, and some others.

We have also considered the effects of dust radial mixing. The radial mixing is modeled as diffusion, using the Schmidt number
from \citet{2007Icar..192..588Y}, i.e.  $D_{\rm dust} = D_{\rm gas}/(1+{\rm St}^2)$. The dust diffusivity is taken to be the
dust viscosity ($D_{\rm gas} = \nu_{\rm gas}$) which is the \citet{1973A&A....24..337S} viscosity for the given alpha value.
We found that radial mixing does not change the disk physical and chemical structure significantly and leads to the gas/dust
temperature increase by several K at intermediate radii.

%% file: diss.tex
\section{Discussion}

\subsection{Comparison to \cite{Vasyunin2011}}

\begin{figure*}[!ht] 
\centering
\includegraphics[width=0.8\textwidth]{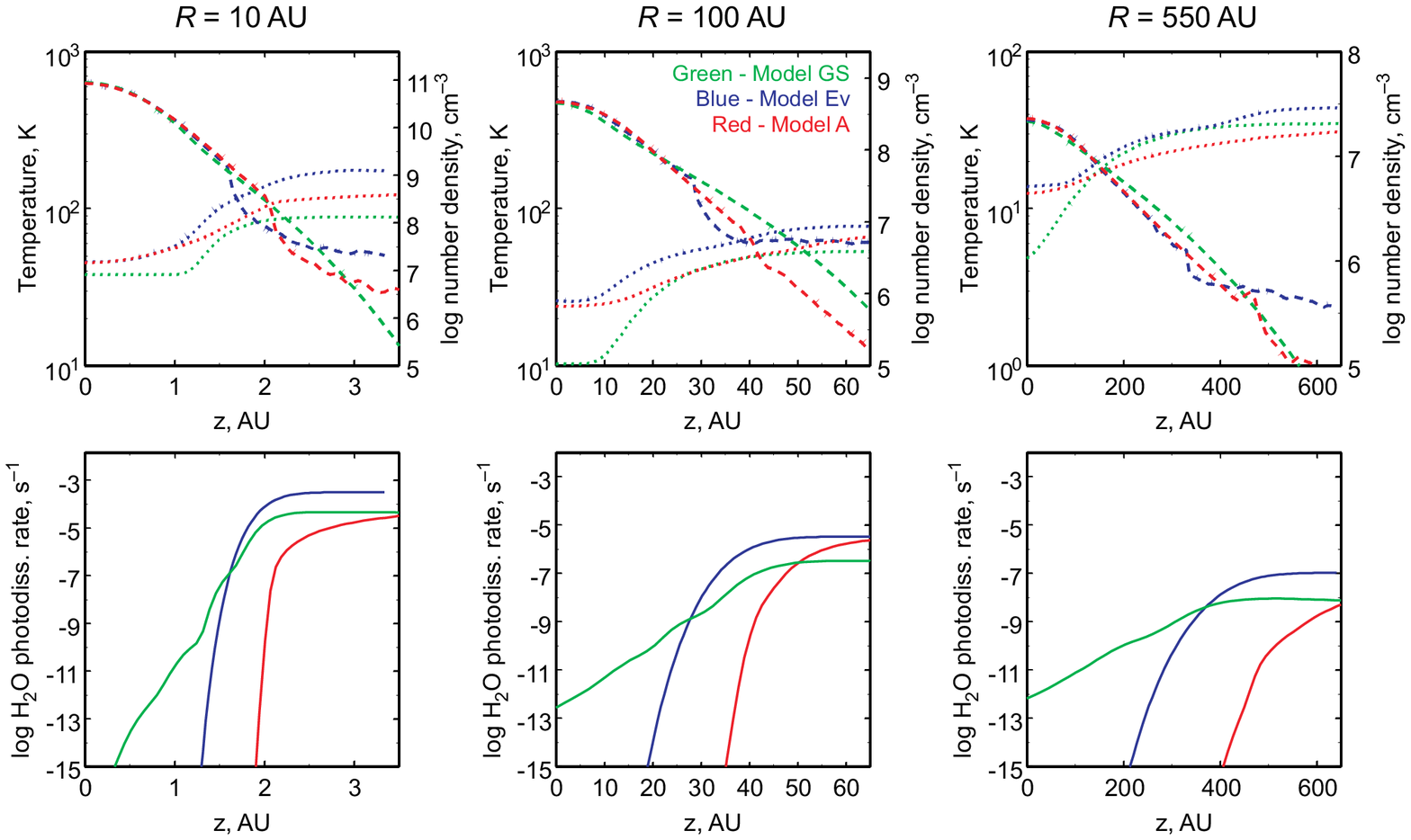} 
\caption{Top row: disk structure in the present study and in \cite{Vasyunin2011}, their Model~GS.
As in Figure~\ref{totstruct}, dashed lines show the gas density, and dotted lines show
dust temperature. Note that dust and gas temperature are equal in Model~GS. Bottom row:
water dissociation rates at the selected radii in the present study and in
\cite{Vasyunin2011}. Different colors correspond to the same models as in the top row.}
\label{oldnewstruct} 
\end{figure*}

While many aspects of the presented model are derived from the model used by
\cite{Vasyunin2011}, the new treatment of the disk  structure results in
parameters that are too different to allow a direct comparison of the `old' and `new'
results. While density profiles are nearly the same in both studies, there are two key
differences in  the disk dust temperature and in  the disk radiation field
(Figure~\ref{oldnewstruct}). First, the improved radiation transfer model makes dust in
the `new' disk midplane significantly warmer than dust in the `old' disk midplane, at
least, in the remote parts of the disk ($R>100$~AU). Second, because of scattering the
`new' disk is less transparent to dissociating far-UV radiation than the `old' one. These
two differences are related to the {\em radiation transfer treatment}, so we may expect
that basic inferences of \cite{Vasyunin2011} on the disk chemical structure should be
retained in the new results, if they are mostly related to the {\em dust evolution\/}.

This is indeed the case, with a few exceptions. First, the general conclusion of
\cite{Vasyunin2011} that dust evolution increases gas-phase column densities of most
species is entirely confirmed in the present study. Second, almost all species, designated
as sensitive to grain evolution in \cite{Vasyunin2011}, like CO, CO$_2$, H$_2$O, C$_2$H,
retain this status in the present study\footnote{A species is marked as sensitive if its
abundances in models with pristine and evolved dust differ by more than an order of
magnitude at least at one of the considered radii.}.

In Table~\ref{tabsens} we show column densities for species listed in Table~2 from
\cite{Vasyunin2011}, along with the newly calculated column densities.
Few remarks are needed. Some species, like methanol, cyanopolyynes or formic
acid, are significantly less abundant in the new model. This is due to generally less
effective surface chemistry in a warmer disk midplane, where depletion of CO and other
similar volatile ices is less severe and where a desorption rate of hydrogen atoms from dust
surfaces is higher. The chances for them to be observed are, thus, slim (within the
framework of our modeling approach). For some species from this list the `sensitivity
region' (the region where the two models differ most) shifts or extends to other radii
(typically, from ten to hundred AU). These are HCNH$^+$ (derived mainly from HCN), NH$_3$,
and OH. Surface hydrogenation also plays an important role in the synthesis of these
species.

Column densities of three species, H$_2$CS, HC$_5$N, and HCO$^+$, while still enhanced by
the dust growth, differ by less than an order of magnitude in the new calculation, so they
do not conform to our sensitivity criterion. Thioformaldehyde that has been mentioned in
\cite{Vasyunin2011} as a molecule most sensitive to dust growth and HCO$^+$ are now
significantly more abundant in the midplane of Model~A due to higher dust temperature and
less severe depletion of their parental species, CO and (H)CS. This shortens the break
between the pristine and evolved dust models.

Abundance of CH$_3$CH$_3$ is also significantly enhanced in the midplane at $R=10$~AU,
relative to results of \cite{Vasyunin2011}, and is nearly the same both in Model~A and in
Model~Ev. As our model has a warmer inner midplane, surface radicals out of which
CH$_3$CH$_3$ is formed become more mobile and reactive. A molecular layer no more
dominates in its column density, so the molecule loses its sensitivity to the dust growth
in the inner disk. In the outer disk the situation is more complicated. There,
CH$_3$CH$_3$ is still sensitive to dust growth, but the sign of the sensitivity is
different. While in \cite{Vasyunin2011} its column density was greater in the model with
pristine dust, now CH$_3$CH$_3$ shares the common behavior and is enhanced in the
molecular layer of Model~Ev due to higher density there.

Another molecule that shows the `reversed' sensitivity is HCN. As we mentioned above, its
abundance in the midplane is higher in Model~A because of more effective surface
synthesis. It also exceeds HCN column density in Model~A5 from \cite{Vasyunin2011} due to
somewhat higher dust temperature, that also intensifies HCN ice production (as surface
production of CN is faster in the warmer ANDES model). At larger radii, HCN behavior is
similar in both studies. These findings demonstrate the importance of the correct treatment of the radiation transport
and also imply that the stellar and interstellar radiation fields need to be discretized
as good as possible.

%\begin{landscape}

\begin{table*} 
\caption{Species sensitive to grain evolution. Observed column densities are
compiled from \cite{Pietu_ea07}, \cite{Dutrey_ea07}, \cite{Chapillon2012},
\cite{Bergin_ea10}, and \cite{Henning_ea10}.} 
\label{tabsens} \centering
\begin{tabular}{l|cccccc|c} 
\hline 
\hline 
Species  & \multicolumn{6}{|c|}{Column densities, cm$^{-2}$} & Observed column \\ 
& \multicolumn{2}{c}{10 AU} & \multicolumn{2}{c}{100 AU} & \multicolumn{2}{c|}{550 AU} & densities, cm$^{-2}$\\ &
Pristine & Evolved & Pristine & Evolved & Pristine & Evolved & DM Tau  \\ 
\hline 
CO       & 2.0(17) 6.5(17)& 1.1(18) 1.8(18)& 2.0(17) 1.1(17)& 9.4(17) 4.3(17)& 1.7(17) 1.3(16)& 2.9(17) 1.7(17)& 1.4(17) \\ 
CO$_2$    & 6.8(10) 6.9(13)& 9.5(13) 7.3(14)& 2.4(12) 1.0(11)& 8.5(14) 6.1(13)& 8.2(13) 1.5(13)& 9.8(14) 7.0(14)& - \\ 
CH$_3$OH    & 1.1(03) 5.8(06)& 3.5(08) 2.6(08)& 1.9(08) 3.8(05)& 1.0(10) 4.6(07)& 4.6(09) 1.2(07)& 1.4(10) 2.0(08)& -  \\
H$_2$O    & 2.5(13) 1.3(12)& 1.1(16) 2.3(16)& 3.6(14) 7.1(13)& 7.0(15) 5.2(16)& 1.4(14) 8.2(13)& 2.0(15) 6.4(16)& $<3.0(13)$  \\ 
H$_2$S    & 1.7(05) 2.9(06)& 1.3(10) 7.8(06)& 2.0(09) 1.2(06)& 1.1(11) 5.1(07)& 2.9(10) 6.3(09)& 3.6(10) 5.0(11)& -  \\ 
C$_2$H    & 6.3(10) 1.8(11)& 2.0(12) 1.0(13)& 2.6(12) 3.0(11)& 2.2(12) 2.9(12)& 6.7(12) 5.7(12)& 5.0(12) 5.1(12)& 2.8(13) \\ 
C$_2$H$_2$   & 6.2(10) 3.1(12)& 4.6(12) 1.2(13)& 4.5(12) 3.9(11)& 3.0(12) 6.1(12)& 4.4(12) 3.0(12)& 1.6(12) 8.3(12)& -  \\ 
CH$_3$CH$_3$ & 3.0(09) 1.1(14)& 1.5(12) 6.9(14)& 1.8(10) 1.6(08)& 3.3(08) 4.1(11)& 1.3(07) 5.1(05)& 2.5(06) 8.4(07)& -  \\ 
H$_2$CS & 8.0(05) 6.4(09)& 3.5(11) 1.8(10)& 2.7(11) 5.3(06)& 8.3(11) 1.5(07)& 2.7(10) 2.2(07)& 1.8(11) 2.2(07)& - \\ 
HCN & 2.6(12) 8.7(14)& 5.0(13) 4.5(13)& 6.9(12) 2.5(11)& 4.2(13) 1.5(13)& 9.0(12) 2.0(12)& 2.1(13) 2.5(13)& 6.5(12)  \\ 
HC$_3$N & 6.8(08) 6.8(07)& 1.7(12) 7.5(11)& 4.1(11) 5.4(08)& 2.8(11) 2.8(11)& 1.1(12) 8.9(10)& 4.3(10) 6.6(11)& - \\ 
HC$_5$N & 4.9(08) 6.7(06)& 1.5(12) 1.9(07)& 2.1(11) 9.9(05)& 1.5(11) 1.0(06)& 1.8(11) 1.6(07)& 8.8(10) 2.4(07)& -  \\ 
HC$_7$N & 4.7(06) 5.8(02)& 1.2(12) 2.1(06)& 6.1(10) 1.6(04)& 5.3(10) 8.5(04)& 9.2(10) 2.2(06)& 2.8(10) 9.8(06)& -  \\
HCNH$^+$ & 2.0(10) 2.1(11)& 2.7(11) 2.1(11)& 5.7(10) 2.1(09)& 1.3(11) 4.6(10)& 6.6(10) 2.5(10)& 6.8(10) 8.6(10)& -  \\ 
HCOOH & 1.4(11) 6.6(06)& 6.3(13) 4.5(08)& 1.1(12) 2.5(07)& 1.3(13) 3.2(08)& 2.6(11) 1.3(09)& 2.1(12) 1.3(09)& -  \\ 
OCN & 3.2(07) 3.0(06)& 4.3(09) 5.5(07)& 2.3(10) 4.5(09)& 1.5(12) 4.2(09)& 1.3(11) 1.6(11)& 4.2(12) 4.3(11)& -  \\ 
OCS & 1.6(07) 1.9(06)& 9.3(10) 9.6(10)& 1.6(10) 3.9(08)& 1.5(10) 9.5(10)& 2.7(08) 1.4(08)& 4.5(08) 3.9(09)& -\\ 
NH$_3$ & 1.7(11) 3.7(10)& 3.2(13) 7.3(12)& 1.1(13) 9.7(11)& 3.5(13) 1.8(13)& 8.5(12) 4.5(12)& 1.2(13) 2.3(13)& -  \\ 
HCO$^+$ & 1.1(11) 2.7(12)& 5.2(12) 8.1(12)& 2.6(12) 4.6(12)& 2.2(12) 1.8(12)& 2.2(12) 7.9(11)& 1.9(12) 1.3(12)& 6.5(12) \\ 
OH & 6.1(13) 1.9(13)& 1.9(14) 2.1(14)& 2.7(13) 1.0(13)& 1.2(14) 9.8(13)& 1.9(13) 9.6(12)& 1.5(14) 2.2(14)& -  \\ 
\hline 
\hline 
\end{tabular} 
\end{table*}

%\end{landscape}

\begin{figure} 
\includegraphics[width=0.5\textwidth]{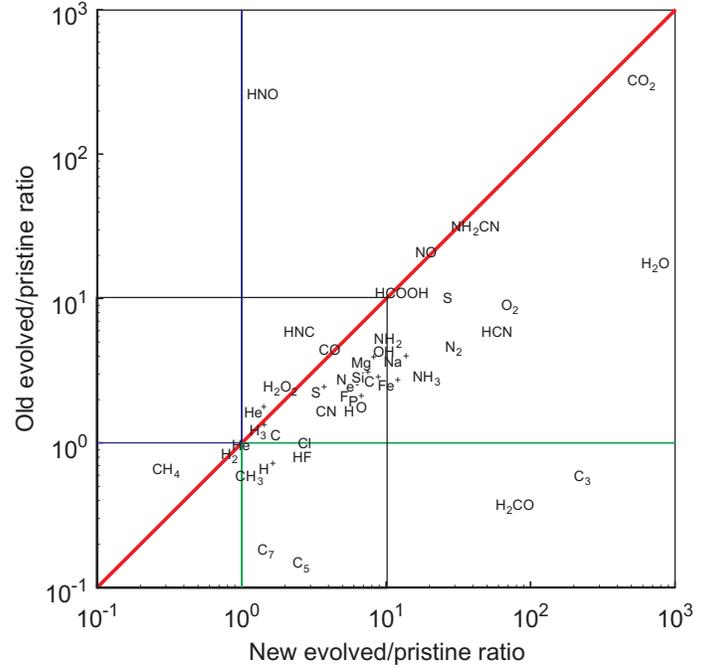}
\caption{Relation between molecular dust growth sensitivity in the present study and in
\cite{Vasyunin2011}. Plotted are ratios of column densities at $R=100$~AU in  the models with
evolved and pristine dust. The red line corresponds to equal sensitivity. Species in upper
right and lower left quadrants show the same kind of sensitivity in both studies. Species
within green lines have their column densities decreased by the dust growth in the
previous study, while in the present study dust evolution increases their column
densities. A black square indicates difference in column densities less than an order of
magnitude, which we interpret as an absence of strong sensitivity.} 
\label{oldnew}
\end{figure}

Table~\ref{tabsens} contains only a few representative species. To have a broader
perspective, we perform a general comparison relating  the column density ratios in  the models
with pristine and evolved dust computed in \cite{Vasyunin2011} and in the present study,
for all species at  10, 100, and 550~AU. Results of comparison for $R=100$~AU are shown in
Figure~\ref{oldnew}. Only gas-phase species with mean abundances greater than $10^{-10}$
are shown. Most species are concentrated around a red line that corresponds to equal old
and new ratios. This indicates that most column densities respond similarly to dust growth
both in \cite{Vasyunin2011} and in the present study. Also, most species reside in the
upper right quadrant, showing that in both studies dust evolution, as a rule, increases
molecular column densities.

Carbon dioxide is most sensitive to dust growth and is, thus, located in the upper right
corner. This is not surprising as its production mostly occurs on grain surfaces via
slightly endothermic reactions of CO and OH. A quite high water sensitivity was obtained in
\cite{Vasyunin2011}, and now it becomes even higher. Similar to water and carbon dioxide,
HNO was very sensitive to dust growth in our old computation because its main production
route is surface synthesis. In the new computation this route is less effective due to
warmer dust, so HNO is mostly produced in the gas phase. This makes it less susceptible to
changes in dust properties. It must be kept in mind that warmer dust has dual effect on
surface chemistry. On one hand, a larger temperature implies more rapid hopping and larger
reaction rates. On the other hand, more volatile reactants evaporate faster from warmer
grains, thus, quenching the formation of some species.

An opposite example is represented by formaldehyde. This species was barely sensitive to
dust growth in the old computation, with column density being slightly smaller in the
model with evolved dust. In the present study, H$_2$CO column density is significantly
greater in the model with evolved dust. This behavior is related to details of  the UV
penetration. In the old models, where only  the UV absorption has been taken into account, the
UV field intensity falls off quite slowly as we go deeper into the disk. Because of that
abundance maxima in the molecular layer for molecules that are mostly susceptible to
photodesorption and photodissociation are extended and shallow. Thus, their column
densities are less sensitive to dust evolution.
Detailed treatment of the radiation transfer in the new model predicts a sharper transition
from the illuminated atmosphere to the dark interior. The molecular layer becomes
significantly narrower and is, thus, much more sensitive to the extent of dust growth and
sedimentation.

The overall conclusion from the presented comparison is the following. We confirm that
dust evolution changes column densities of many molecules (see Table~\ref{tabsens} and
Figure~\ref{oldnew}). Most species that have been listed as especially sensitive to dust
evolution in \cite{Vasyunin2011} retain this status in the present study. However, column
densities of some species turn out to depend on the details of the radiation transfer
treatment, and this dependence will become even stronger when we will proceed from column
densities to line intensities. ANDES makes all the necessary preparatory work for that,
providing us with both abundances and gas temperatures.

However, there is another aspect, apart from the radiation transfer, that may affect our
conclusions. This aspect is related to possible evolutionary changes. As in ANDES we use
time-dependent chemistry, we can  provisionally estimate its importance.

%% file: timechem.tex
\subsection{Disk structure with time-dependent chemistry} 

\begin{figure*}[!ht] 
\includegraphics[ scale=.33, angle=270]{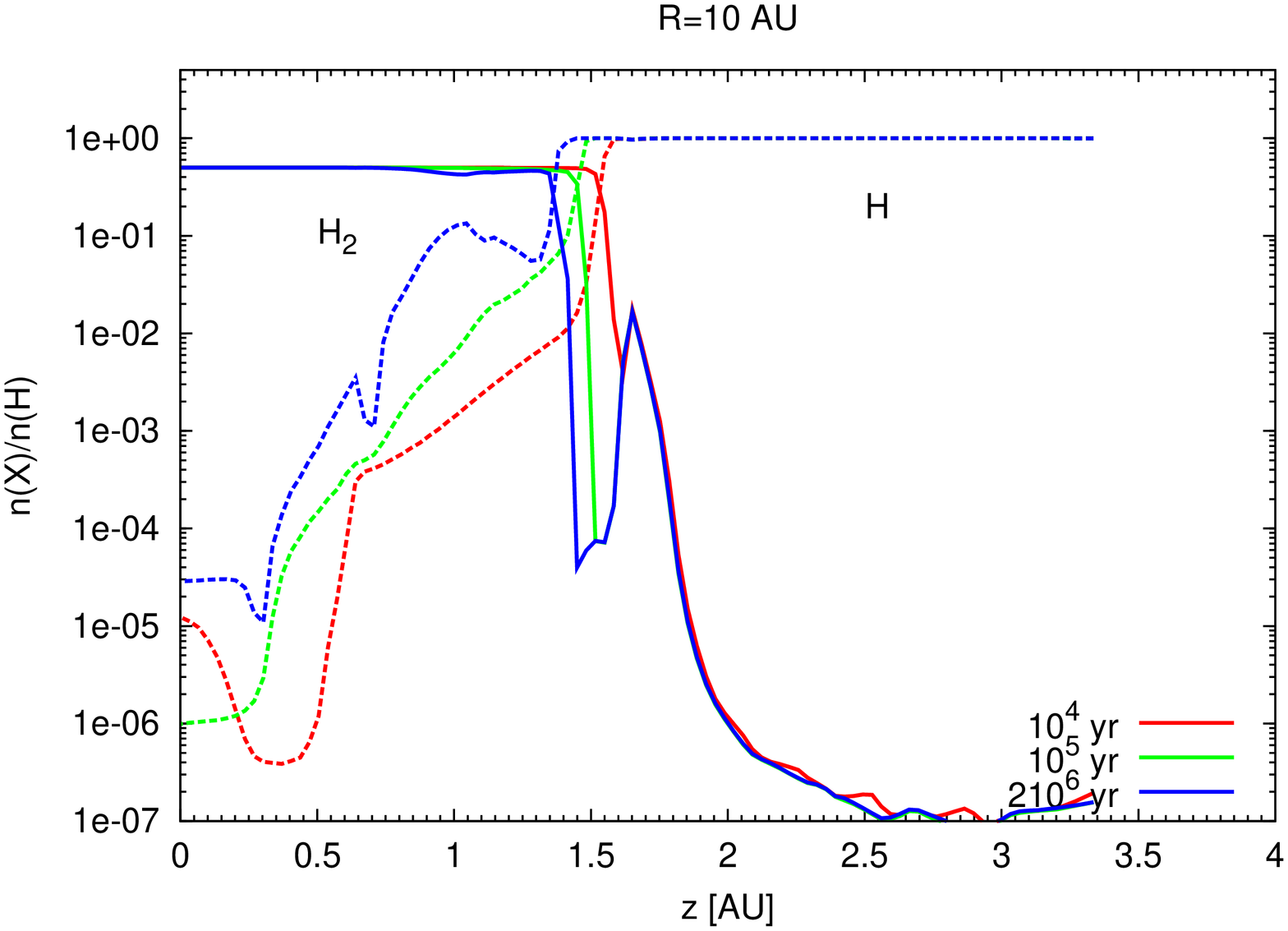}
\includegraphics[scale=.33, angle=270]{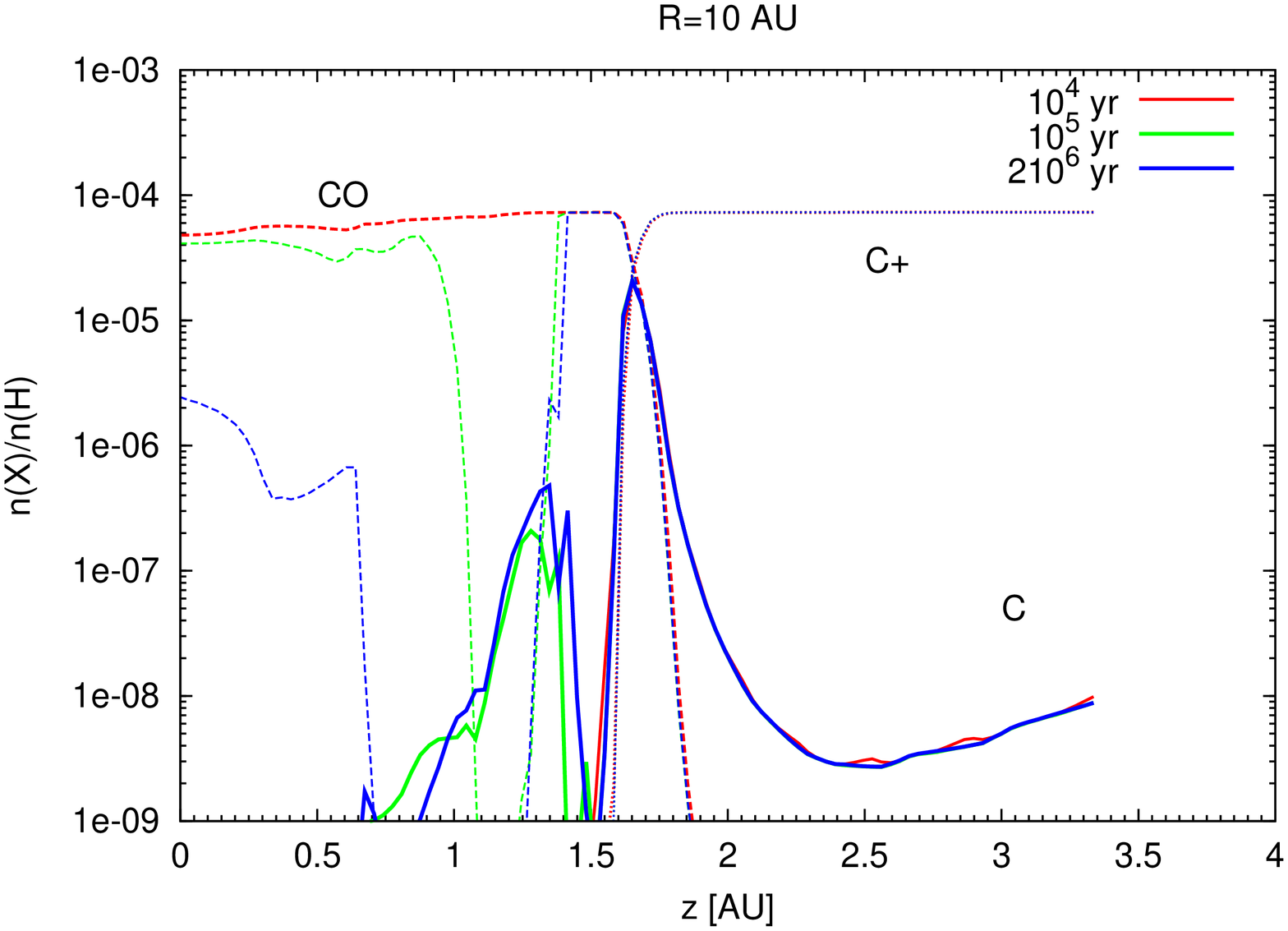} \\
\includegraphics[ scale=.33, angle=270]{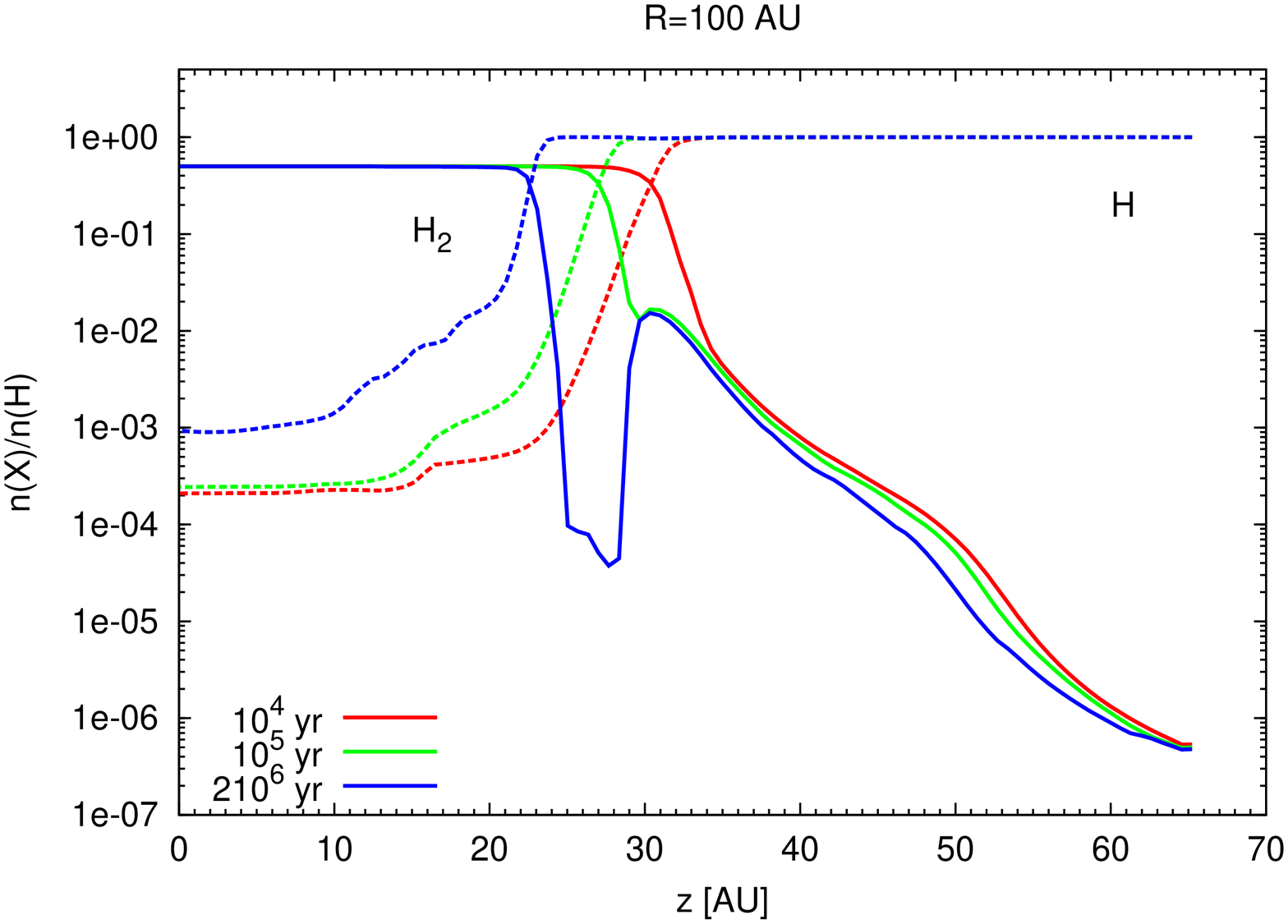}
\includegraphics[scale=.33, angle=270]{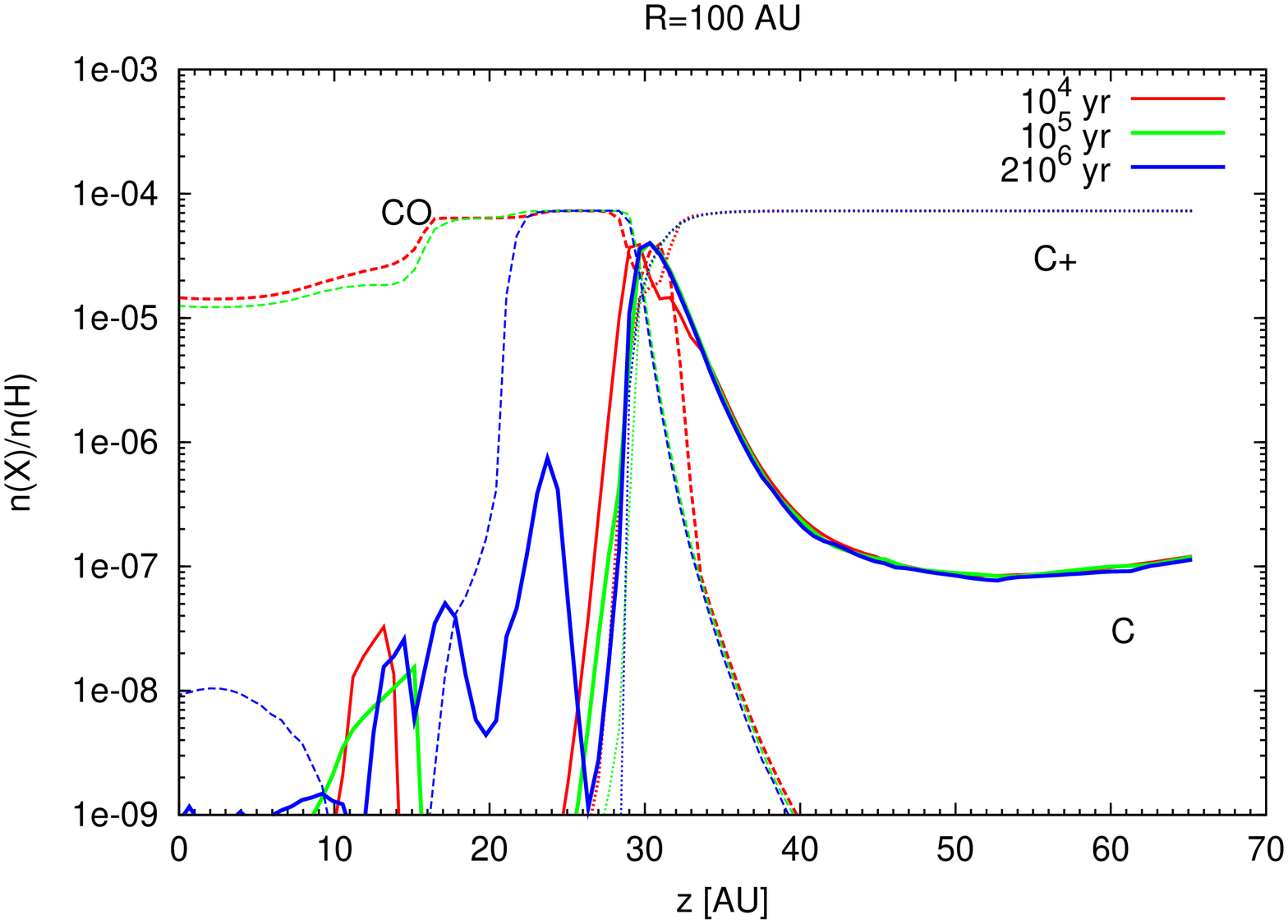} \\
\includegraphics[ scale=.33, angle=270]{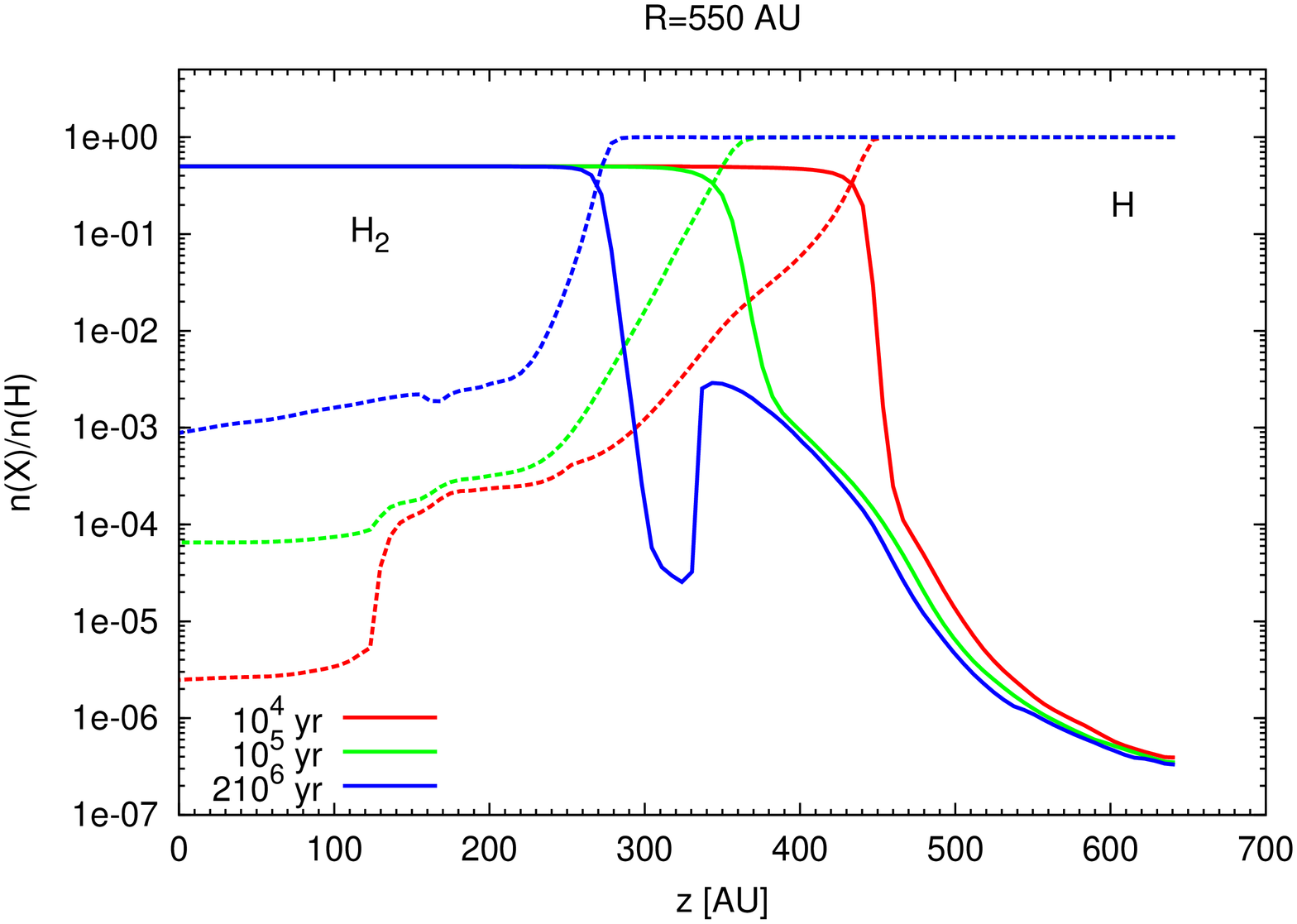}
\includegraphics[ scale=.33, angle=270]{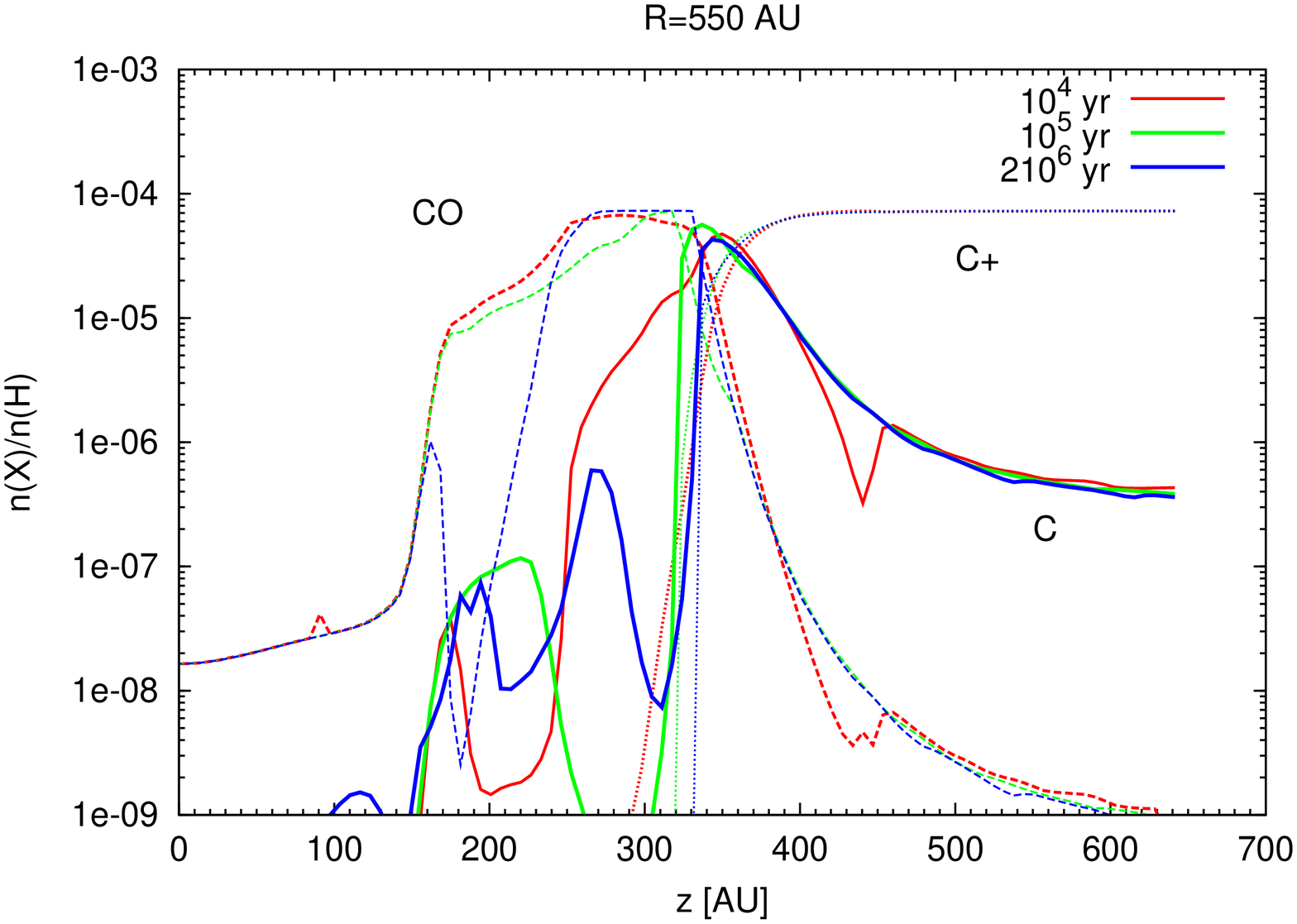}
\caption{The left panels show \HH/H transition at selected radii calculated for different epochs: $10^4$,  $10^5$, and $2\cdot
10^6$~yrs for the disk with evolved dust. The right panels show the CO/C/C$^+$ transition.} 
\label{chem_epoch} 
\end{figure*}

In order to demonstrate the effect of time-dependent chemistry on the
disk  chemical structure we perform model calculations with abundances of major molecular coolants at $10^4$,  $10^5$, and $2\times
10^6$~years. We assume that the disk chemically evolves from mostly neutral atomic gas, with molecular hydrogen and a low fraction
of atomic hydrogen ($10^{-3}$ to the total number of hydrogen nuclei). We do not consider the time evolution of dust grain
distribution in order to focus on effects of chemical evolution, and utilize vertical dust distribution after 2~Myr. Results are
presented in Figure~\ref{chem_epoch}, showing the relative abundances of H$_2$, H, CO, C, and C$^+$ as a
function of height at the distinct epochs for disk radii of 10, 100 and 550~AU.

As can be clearly seen, the location of the H$_2$/H transition shifts toward the midplane with time for all the considered radii
due to slow photodissociation of molecular hydrogen, self-shielded from the strong FUV stellar radiation. H$_2$ cannot be quickly
re-formed in this region in  Model~Ev due to overall lack of grains, providing catalytic surface for H + H reaction.
Consequently, between $10^4$ and $2\,10^6$~years, at radii of 10, 100, and 550~AU, the PDR zone shifts from 1.6 to $\approx
1.4$~AU, from 31 to 22~AU, and from 425 till 290~AU, respectively. This effect is more pronounced in the outer disk, where
densities and density gradient are lower.

An interesting feature of the chemical structure in Model Ev is the presence of a `dip' in H$_2$ vertical distribution at the final time moment
at all radii. This region with depression in H$_2$ concentration is caused by its slow X-ray/UV destruction, which cannot be
compensated by the H$_2$ surface production on a few remaining grains. However, just above this depression region dust-to-gas
ratio locally increases, and so does the available surface for hydrogen recombination (per unit gas volume). The reason  for the
elevated dust-to-gas ratio is a  gas redistribution from the top of the coupling region to greater heights due to extra heating.

In contrast, the evolution of ionized carbon reaches a chemical steady-state rapidly everywhere in the disk thanks to its fast
ion-molecule chemistry pathways, so the C$^+$ concentration is not time-sensitive (after $10^4$~years), see
Figure~\ref{chem_epoch}. The chemical evolution of C$^+$ is governed by a simple and limited gas-phase
reaction network in the disk atmosphere, where it is an important coolant with a relative abundance of $\approx 10^{-4}$
\citep[see, e.g.][]{2004A&A...417...93S,2011ApJS..196...25S}. Neutral atomic carbon shows little time evolution, if any, in the
disk regions adjacent to the midplane at radii smaller than $\sim 100$~AU. Due to relatively large densities in disk equatorial
regions, neutral carbon is rapidly converted to CO and hydrocarbons. This is not true for lower-density outer disk regions, at
$R\ga500$~AU and $z/R \sim 0.3-0.6$, where the C abundance changes substantially with time. Since initially all elemental carbon
is in the atomic form, in the outer disk, less dense and more transparent to the interstellar FUV radiation, conversion of C into
CO and hydrocarbons takes more than $10^4$~years (see Figure~\ref{chem_epoch}, bottom row).

The gas-phase CO abundances follow the pattern of H$_2$/H and C and do not reach a steady-state within 2~Myr everywhere in the
disk model with evolved dust.
The grain growth increases the CO freeze-out timescale to $\ga 1$~Myr  in the inner and intermediate radii (Figure~\ref{chem_epoch},
top and middle rows).  In the midplane, where gas-phase CO abundance is low, this molecule is present as CO ice. The final distribution of the CO
abundance at $R\la 100$~AU shows an interesting feature: due to severe grain growth CO freeze-out is inefficient in the midplane,
but still effective at disk heights of $z/R \approx 0.2$~AU. At even larger heights the CO molecular layer starts, so CO
emission lines are excited both in the very cold and warm regions. Since $^{12}$CO, $^{13}$CO and C$^{18}$O isotopologue lines,
having vastly different opacities, allow probing these two temperature zones, this should be visible with modern
radio-interferometers. Intriguingly, evidence for the presence of very cold CO gas was found by \citet{Dartois_etal2003}, and,
later, for other molecules like HCO$^+$, CCH, CN, and HCN \citep[see discussion in][]{2011ApJS..196...25S}.

\begin{figure}[!ht] 
\begin{center} 
\includegraphics[ scale=.35, angle=270]{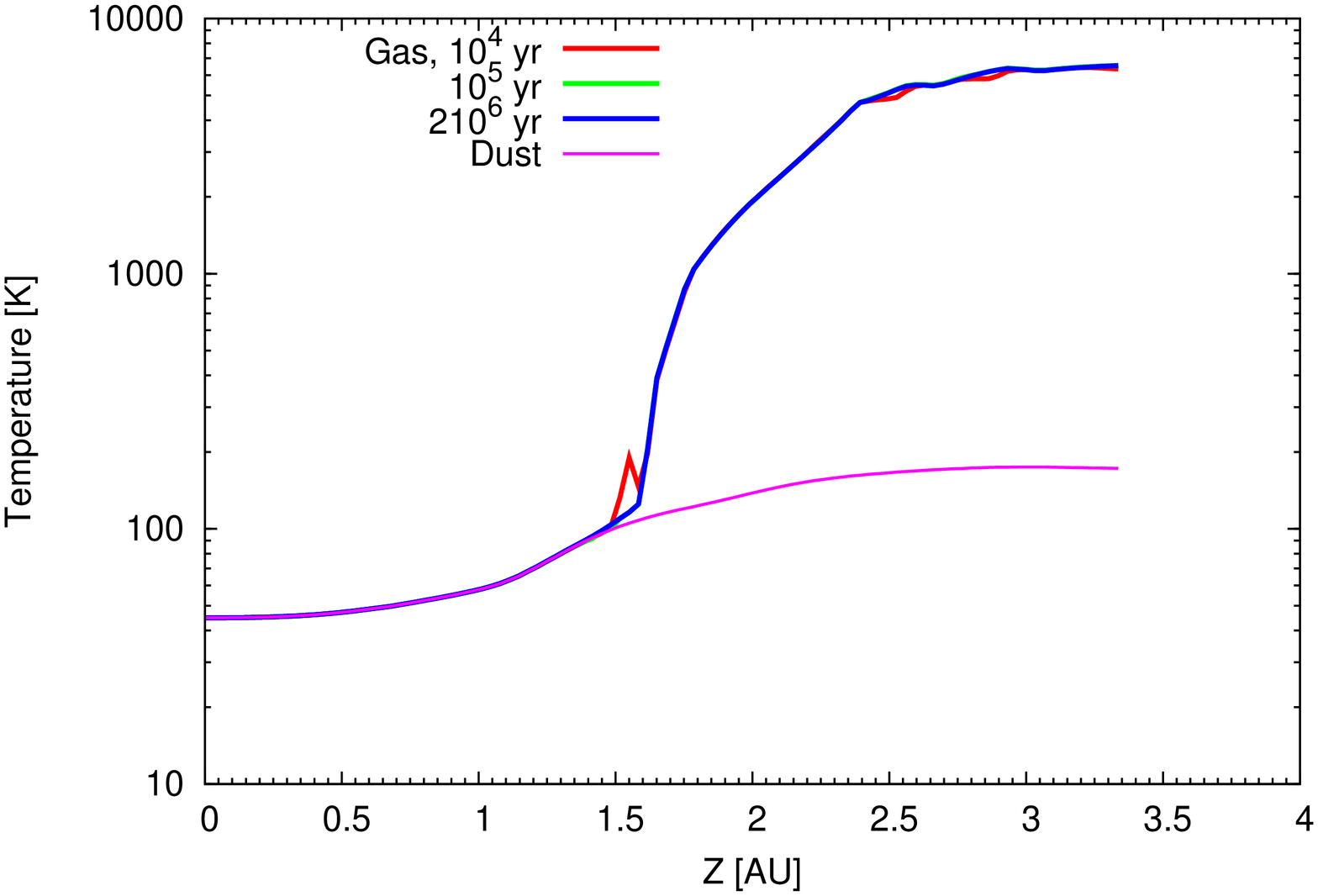}\\
\includegraphics[ scale=.33, angle=270]{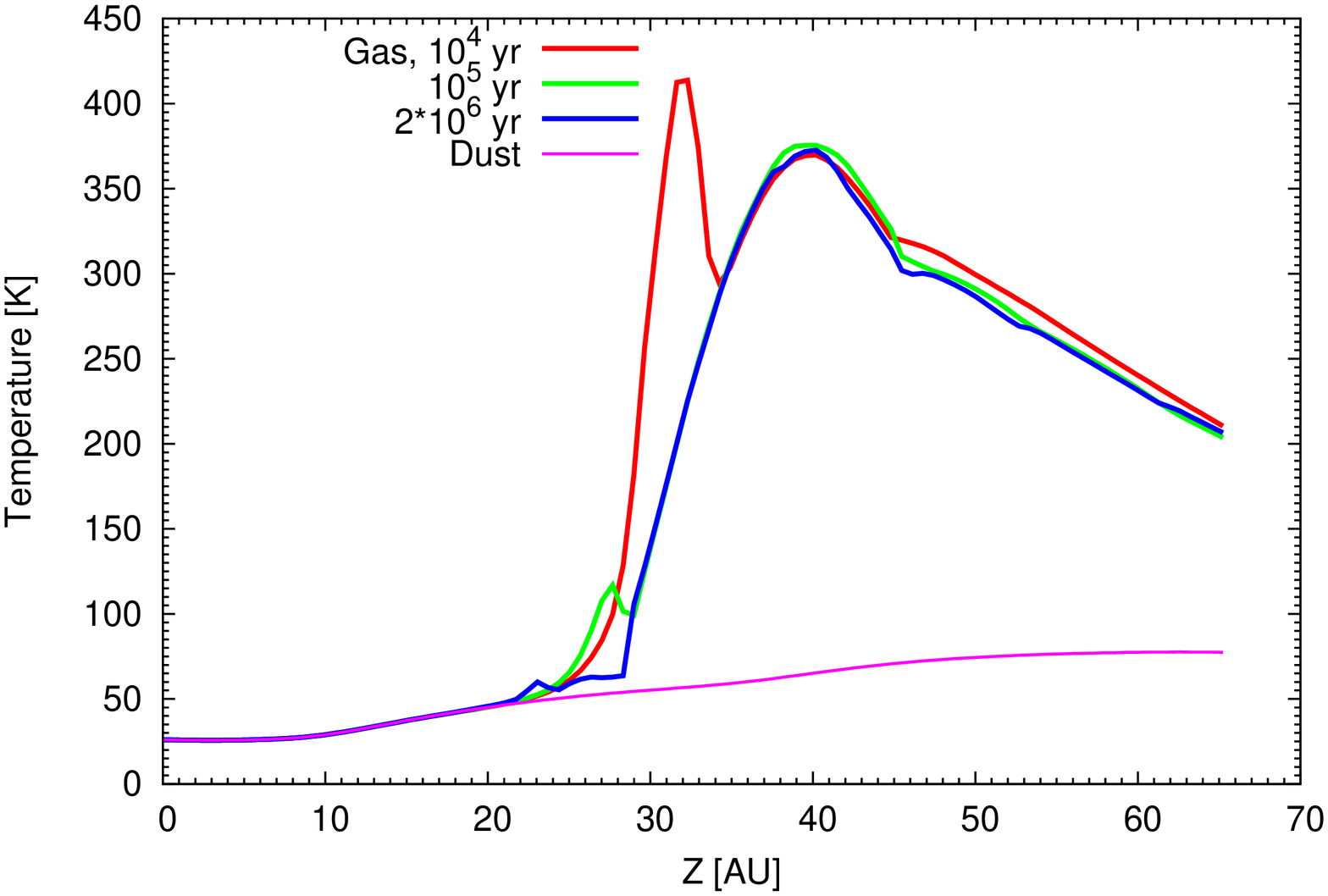}\\
\includegraphics[ scale=.33, angle=270]{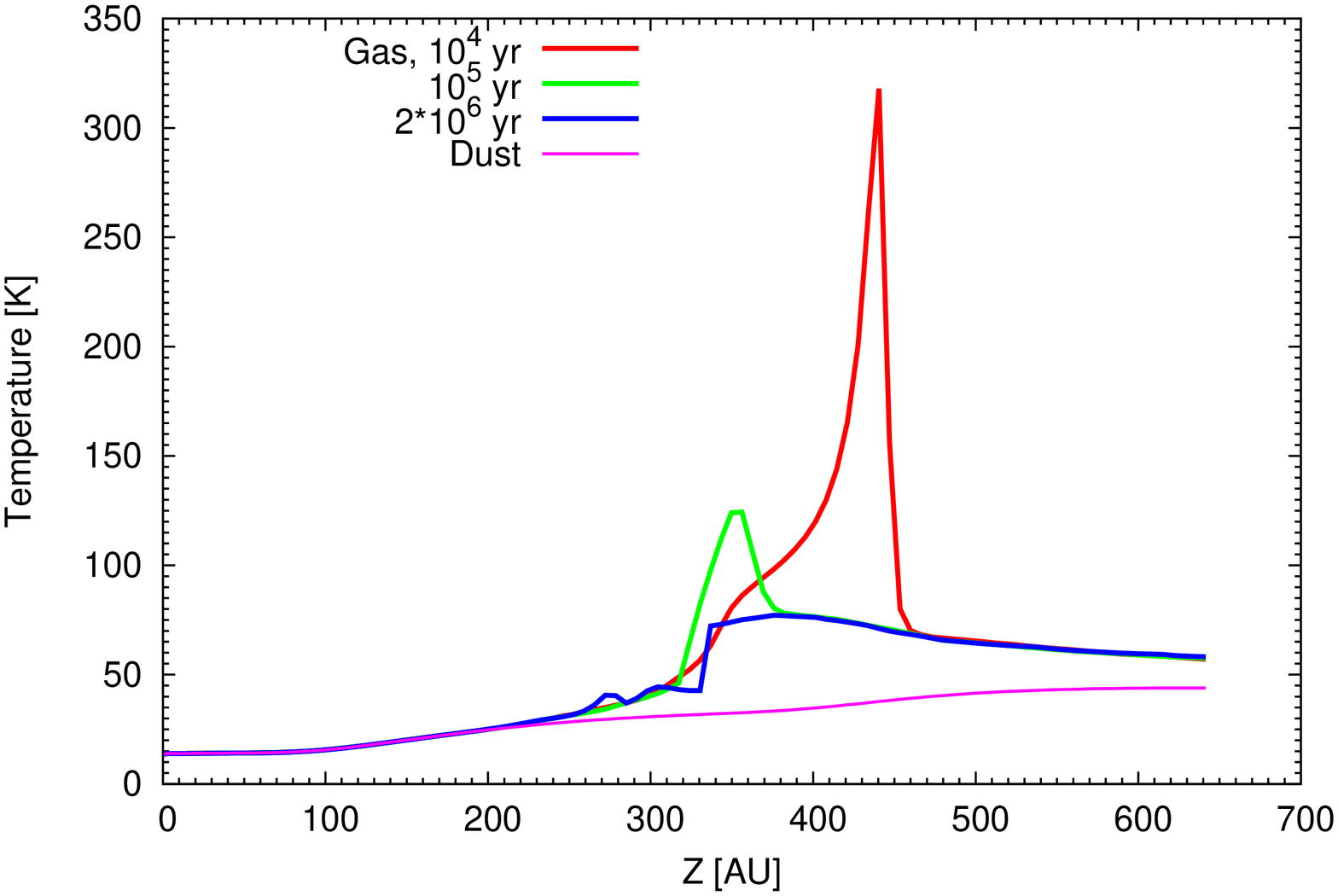} 
\caption{Gas temperature profiles at selected radii (10, 100, 550~AU from top to bottom) calculated with chemical abundances at different epochs: $10^4$,  $10^5$, and
$2\cdot 10^6$~yrs.} 
\label{temp} 
\end{center} 
\end{figure}

Enhanced amounts of H$_2$ at $10^4$~years in disk regions with high FUV radiation intensities lead to additional heating by the
UV-excited H$_2$. Since the H$_2$/H boundary is moving down, the gas thermal structure of the disk responds accordingly and also
shows strong variations of $T_{\rm g}$ in a narrow disk layer, in particular, at $R>100$~AU (see
Figure~\ref{temp}). While the gas temperature varies from 250~K to $\approx 200$~K (25\%) at $R=10$~AU, at the outer
disk region, $R=550$~AU, the gas temperature difference at various times is about 250~K (from 320~K to $\approx 75$~K, or a factor
of 4), compare top and bottom panels of Figure~\ref{temp}. Naturally, it should also have a strong impact on chemical composition
and appearance of the disk molecular layer, from which most of line emission emerges. More importantly, it demonstrates the
importance of using the time-dependent chemistry for calculating abundances of key gaseous coolant instead of the commonly applied
steady-state approach.

%---------------------------------------------------------------

%% file: conclusions.tex
\section{Conclusions} 
A multi-dimensional self-consistent model of protoplanetary disks
`ANDES' is introduced and described. The purpose of ANDES is to provide researchers with a
state-of-the-art, most up-to-date detailed thermo-chemical model of a protoplanetary disk
that can be used to analyze high-quality IR and (sub-)millimeter observations of
individual nearby disks. For the first time grain evolution and large-scale time-dependent
molecular chemistry are included in modeling of physical structure of protoplanetary
disks.

The iterative ANDES code is based on a flexible modular structure that includes 1) a 1+1D
continuum radiative transfer module to calculate dust temperature, 2) a module to
calculate gas-grain chemical evolution, 3) a 1+1D module to calculate detailed gas energy
balance, and 4) a  1+1D module that calculates dust grain evolution. The disk structure is
computed iteratively, assuming fixed dust density structure after the first iteration.
Typically it takes $\sim 10$ iterations to reach convergence at 1\% level of accuracy.

The continuum radiative transfer module is based on the two-stream Feautrier method with
a high-resolution frequency grid. We consider dust continuum absorption, thermal emission,
and coherent isotropic scattering. The dust evolution is modeled by accounting for
coagulation, fragmentation, and gravitational sedimentation towards the disk midplane
balanced by turbulent upward stirring. The chemical model is based on a
gas-grain realization of the RATE'06 network, and includes surface reactions and X-ray/UV
processes. All modules have been thoroughly benchmarked with previous studies, with
overall good agreement and performance.

We study the impact of dust evolution on dust temperature, gas temperature, and chemical
composition by comparing results of the disk models with evolved and pristine dust. We
compute gas thermal structure corresponding to chemical abundances evolving from the
initial abundances for $10^4$,  $10^5$, and $2\cdot 10^6$~years. We show that
time-dependent chemistry is important for a proper description of gas thermal balance. The
strongest impact on the gas temperature (up to 100~K) occurs in the outer, low-density
region beyond 100~AU. This is mainly due to the shift of H$_2$/H PDR transition deeper
into the disk with time.

In accordance with previous studies, it is found that the gas becomes hotter than the dust
in elevated disk regions reaching 1\,000--10\,000~K in the inner atmosphere. However, the
main heating source is different for the two dust models. In the disk with pristine dust it is
photoelectric heating by grains. In the atmosphere of disk with evolved dust grains are
strongly depleted, therefore photoelectric heating by PAHs becomes a dominant heating
process. Thus a realistic, observationally-based estimates of absolute PAH abundances and
sizes are required to calculate accurately gas temperature in the inner, $\sim 1-50$~AU
disk atmosphere accessible with Spitzer, Herschel, and ALMA.

The response of disk chemical structure to the dust growth and sedimentation is twofold.
First, due to higher transparency a partly UV-shielded molecular layer is shifted closer
to the dense midplane. Second, the presence of big grains in the disk midplane delays the
freeze-out of volatile gas-phase species such as CO there, while in adjacent upper layers
the depletion is still effective. Even though the dust evolution shifts the molecular layer of the
water vapor closer toward the cooler, midplane disk region, it
increases its overall concentration. This aggravates the disagreement
between the water vapor column densities predicted by modern
astrochemical models, which are higher than those observed with
Herschel in the disks around TW Hya \citep{2011Sci...334..338H}
and DM Tau \citep{Bergin_ea10} by factors of at least several
(see also discussion in \citet{2011ApJS..196...25S}). Overall, molecular concentrations and thus column
densities of many species are enhanced in the disk model with dust evolution, e.g.,
CO$_2$, NH$_2$CN, HNO, H$_2$O, HCOOH, HCN, CO.

%% file: acknowledgements.tex
\section*{Acknowledgements} 
This research made use of NASA's Astrophysics Data System.
This work is supported by the RFBR grants 10-02-00612, 12-02-31248, Federal Targeted
Program ``Scientific resources of Innovation-Driven Russia'' for 2009-2013 and
NSh-3602.2012.2. SZ is supported by the {\it Deutsche Forschungsgemeinschaft} through
SPP~1573: ``Physics of the Interstellar Medium". DS acknowledges support by the {\it
Deutsche Forschungsgemeinschaft} through SPP~1385: ``The first ten million years of the
solar system~--- a planetary materials approach'' (SE 1962/1-1 and 1-2). A. V. acknowledges the support of the National Science Foundation (US) for the astrochemistry program at the University of Virginia. We thank Kees
Dullemond, Andras Zsom, Ewine F. van Dishoeck and Simon Bruderer for valuable discussions. We highly appreciate comments and suggestions of an anonymous referee, that helped us a lot to improve the quality of the paper.

%% file: appendix.tex
\appendix
\section{Main gas heating and cooling processes}
\label{AppendA}
\subsection{Main heating processes} 
\paragraph*{Photoelectric heating by grains} We follow \cite{2001A&A...373..641K} and
calculate the photoelectric heating rate by silicate grains as 
\begin{equation} 
\Gamma_{\rm PE} = 2.5 \times 10^{-4} k_{\rm abs}^{\rm UV} \epsilon_{\rm dust} \chi, 
\label{GPE}
\end{equation} 
where $k_{\rm abs}^{\rm UV}$ is the dust absorption coefficient at UV
wavelengths,  $\chi$ is the strength of the UV radiation field in units of the Draine FUV
interstellar field \citep{Draine1978}, and $\epsilon_{\rm dust}$ is the photoelectric
efficiency determined by the grain charge parameter $x = \sqrt{T_{\rm g}} {\chi}/{n_{\rm
e}}$ (here $n_{\rm e}$ is the electron number density). For $\epsilon$ we adopt
expressions from  \cite{2001A&A...373..641K}.  The relative strength of the FUV, $\chi$, is
defined as 
\begin{equation} 
\chi =\frac{\int_{\rm 91.2nm}^{\rm 110nm} \lambda u_\lambda d\lambda}{\int_{\rm 91.2nm}^{\rm 110nm} \lambda u_\lambda^{\rm Draine} d\lambda}.
\end{equation} 
 The average dust opacity at UV wavelengths is determined by integration
of frequency-dependent dust absorption cross-sections in the  UV frequency range
\begin{equation}
k_{\rm abs}^{\rm UV} = \frac{1}{\Delta \nu} \int \int f(a) \pi a^2 Q_{\rm abs}(a) da  d\nu,
\end{equation}
where $f(a)$ is given by the dust evolution model.
%--------------------------- 
\paragraph*{Photoelectric heating by PAHs} 
Polycyclic aromatic hydrocarbons (PAHs) possess large cross-sections for UV photon absorption and
therefore can efficiently heat gas by photoelectric emission, even if their abundance is low. 
Heating by PAHs can be particularly important for disks with evolved dust, since PAHs are
better mixed with gas than macroscopic dust particles, and thus remain in the disk
atmosphere while  bigger grains settle towards the midplane \citep{2007A&A...473..457D}.
\cite{Bakes:1994p5701} derived a simple analytical expression for
their PE heating rate: 
\begin{equation} 
\Gamma^{\rm PAH}_{\rm PE} = 10^{-24} f_{\rm PAH}  \htot \epsilon_{\rm PAH} \chi, 
\end{equation} 
where $\htot$ is the hydrogen nuclei number
density, and $\epsilon_{\rm PAH}  = 0.0487 / (1+4\times 10^{-3} x^{0.73})$. The parameter
$f_{\rm PAH}$ is the depletion factor of the PAH abundance relative to the diffuse ISM
value, which is estimated to be $\sim 10-20\%$ of the total carbon budget \citep{Draine:2007p6602}. The
details of the evolution of PAHs in protoplanetary disks are far from being fully understood, though it is
clear that high-energy stellar radiation may play an enormous role in their destruction
and chemical transformation \citep{2010ApJ...718..558A,2010A&A...511A...6S,2012arXiv1204.3651S}. Therefore,
we do not consider PAHs in the simulations of dust  evolution and treat $f_{\rm PAH}$ as a
free parameter of the model. In the present paper we assume $f_{\rm PAH}=0.1$ based on
estimates from observations of PAH spectra in disks surrounding young T~Tauri and
Herbig~Ae stars \citep{2008ApJ...684..411K,2011EAS....46..271K}.  A detailed study of  the effects of PAHs 
heating on the structure of protoplanetary disks with
evolved dust is beyond the scope of the present paper.

%---------------------------

\paragraph*{Cosmic ray heating} Cosmic ray (CR) particles deposit energy mainly through ionization of H$_2$
and H at the rate \citep{Bakes:1994p5701}: 
\begin{equation} \Gamma_{\rm CR} = \zeta_{\rm CR} \left(5.5 \times 10^{-12} n({\rm H}) + 2.5\times 10^{-11} n({\rm H_2})\right),
\end{equation} 
where $\zeta_{\rm CR}$~s$^{-1}$ is the  attenuated CR ionization rate and
$n({\rm X})$ denotes concentration of a species X.

%---------------------------

\paragraph*{Heating by surface H$_2$ formation} Formation of one \HH molecule on the grain
surface liberates 4.48~eV of energy, but the exact partitioning of this energy into \HH
vibration, rotation, translation and accommodation by a grain lattice remains uncertain. It
is commonly assumed that this energy is equally redistributed between rotational,
vibrational and translational movements. We assume that formation of one hydrogen
molecule returns only 1.5 eV ($2.4\times 10^{-12}$~erg) to the gas
\citep{1976ApJ...203..132B}. Then, the corresponding heating rate is 
\begin{equation}
\Gamma_{\rm H_2form} = 2.4\times 10^{-12}  R_{\rm H_2 form}  \htot, 
\end{equation} where
$R_{\rm H_2 form}$ is the $\rm H_2$ formation rate in s$^{-1}$. The  further details of calculation
of chemical reactions rates are described in Sect.~\ref{chemistry}.

%---------------------------

\paragraph*{Photodissociation of H$_2$} We take into account only spontaneous radiative dissociation of \HH:
$\HH+h\nu \rightarrow \HH^* \rightarrow  {\rm H}+{\rm H} + h\nu$. Assuming that the
average kinetic energy of dissociation products is 0.45~eV \citep{1973ApJ...186..165S},
the corresponding heating rate is
\begin{equation} \Gamma_{\rm H_2dis} = 6.4
\times 10^{-13} R_{\HH \rm phdis} n({\rm H_2}), 
\end{equation} 
where $R_{\HH \rm phdis}$
is the photodissociation rate of \HH. To calculate this rate, we take into account
self-shielding of H$_2$ molecules as given by Eq.(37) from \citet{DB96}.
%---------------------------

\paragraph*{Collisional de-excitation of H$_2$} In dense PDR regions collisional
de-excitation of FUV-pumped H$^*_2$ is the second most important heating mechanism
\citep{1995ApJS...99..565S}. Here we adopt a simple two-level approximation of \HH
vibrational heating and cooling from \cite{2006A&A...451..917R}, which nevertheless well
reproduces the net heating rate computed by \cite{1995ApJS...99..565S} with 15 vibrational
levels. The net vibrational heating is given by the following expression: 
\begin{equation}
\Gamma_{net}=\Gamma_{\rm H^*_2}  -  \Lambda_{\rm H_2},
\end{equation} 
where $\Gamma_{\rm H^*_2}$  is the vibrational heating rate by collisional de-excitation and $\Lambda_{\rm H_2}$
is the vibrational cooling rate. 
For  details of the calculation of the  heating and cooling rates we refer to the Appendix C in
\cite{2006A&A...451..917R}. 

%---------------------------

\paragraph{C photoionization} Ionization of atomic carbon releases electrons with kinetic
energies of  $\sim 1$~eV \citep{1987ASSL..134..731B}. The corresponding heating rate can
be approximated as: 
\begin{equation} 
\Gamma_{\rm C} = 1.6 \times 10^{-12} R_{\rm Cph} n({\rm C}), 
\end{equation} 
where $R_{\rm Cph}$ is the photoionization rate of  the C atoms.

%---------------------------

\paragraph{Viscous heating} The viscous heating rate is given by
\citep{1992apa..book.....F}: 
\begin{equation} 
\Gamma_{\rm vis} = \frac{9}{4} \rho \nu_{\rm kin} \Omega_{\rm kep}^2, 
\end{equation} 
where the kinematic viscosity of the gas is
parameterized as $\nu_{\rm kin}=\alpha c_T H_{\rm g}$ \citep{1973A&A....24..337S}, $c_T$
is the isothermal sound speed, $H_g$ is the gas pressure scale height, and $\Omega_{\rm
kep}$ is the Keplerian velocity.
%---------------------------------------------------------------

\subsection{Main cooling processes} 

%---------------------------

\paragraph*{NLTE line cooling} The net line cooling rate for a  given species is
determined by the total amount of upwards and downwards radiative transitions. Level populations for each
coolant are calculated from statistical equilibrium equations.
Unlike FUV, the local FIR intensity that enters these equations depends on the temperature and level populations
 over the large part of the disk. This requires iterations
over all vertical grid points simultaneously. To simplify a calculation, we adopt an escape probability approach using the 
expression (B9) in \cite{Tielens:1985p6987}.

%---------------------------

We perform the full non-LTE calculations, considering the major
coolants for a typical PDR: fine structure lines of  C, O, C$^{+}$ and rotational
lines for the CO molecule. The data for energy levels, collision, emission and absorption
coefficients for computation of the NLTE line cooling are taken from the LAMDA database
\citep{Schier:2005ja}. The data include collision rate coefficients for collisions of \HH,
H, e$^-$, He, and H$^+$ with O and C atoms, as well as collisions of \HH, H, and e$^-$ with
C$^+$, and \HH with CO. For minor coolants we use approximate formulas presented below.

%---------------------------

\paragraph*{High-temperature coolants} The cooling by emission from metastable levels of
neutral and ionic species becomes important at temperatures exceeding several thousand
Kelvin. We calculate the cooling rate from $\rm ^1D - ^3P$ emission by O~I (630~nm)
according to \cite{Sternberg:3p8063}: 
\begin{eqnarray} 
\Lambda_{\rm O I 630} = 1.8 \times 10^{-24}  n({\rm O})  n_{\rm e} \exp{(-22800/T_{\rm g)}}, 
\end{eqnarray} 
where $ n_{\rm O}$ is the neutral oxygen concentration.

The cooling by electron impact excitation of metastable levels of ionic species (e.g.,
Fe$^{+}$, Fe$^{++}$, Si$^{+}$) is calculated by approximate formula from 
\cite{1972ARA&A..10..375D}: 
\begin{equation} 
\Lambda_{\rm ion} (T) = A_i T_{\rm g}^{-1/2} \exp{(-T_i/T_{\rm g})}, 
\end{equation} 
where parameters $A_i$ and $T_i$ for each ion are
given in \cite{1972ARA&A..10..375D}.

Another important cooling process at high temperatures is Ly$\alpha$ emission. We adopt
the cooling rate by Ly$\alpha$ emission from \cite{Spitzer1978}: 
\begin{equation}
\Lambda_{Ly\,\alpha} = 7.3 \times 10^{-19}  n_{\rm e}  n({\rm H}) \exp{(-118400/T_{\rm g})}. 
\end{equation}

%---------------------------

\paragraph*{H$_2$O line emission} 
Rotational line emission of  the H$_2$O molecule can
contribute to cooling in dense disk regions. We include line cooling rates of H$_2$O due
to  the excitation by \HH, using analytical fits from \cite{Neufeld:1993p7740} for $T > 100$~K
and from \cite{1995ApJS..100..132N} for 10~K$< T <$100~K.

\paragraph*{Thermal accommodation} Thermal accommodation is the energy exchange by
inelastic collisions between dust and gas. In disk models with standard ISM-like dust it
is a dominant cooling process, with the exception of the upper, tenuous atmosphere and outer radii
($R>400$~AU) \cite[e.g.,][]{2009A&A...501..383W}. We utilize the corresponding cooling
rate from \cite{Burke:1983p6819}:
\begin{equation} 
\label{gasgrain} 
\Lambda_{\rm d-g} = 4\times10^{-12} \pi \left\langle a^2 \right\rangle n_d \htot \alpha_T \sqrt{T_{\rm g}} (T_{\rm g} - T_{\rm d}), 
\end{equation} 
where the thermal accommodation coefficient $\alpha_T$ is set to 0.3 ( a typical value for silicates and carbon).

%---------------------------------------------------------------

\section{Radiative transfer and gas thermal balance benchmarking}
\label{AppendB}
\begin{figure}[!ht]\begin{center}
\includegraphics[scale=0.3,angle=270]{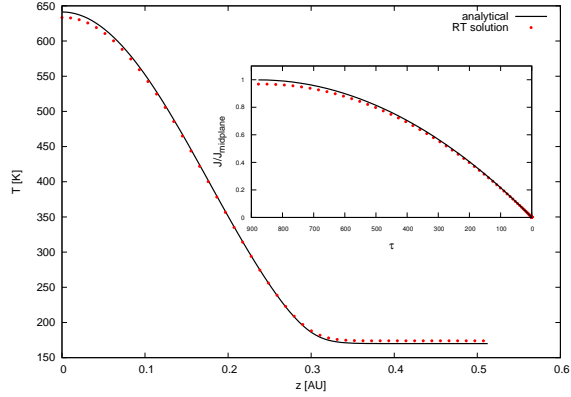} 
\end{center}
 \caption{ Computed and analytical temperatures and mean intensities for a test case.} 
  \label{RTbenchFig}
  \end{figure}

The RT module of the ANDES code was checked for the following cases which allow an analytic or semi-analytic solution:
\begin{itemize}
  \item Optically thin case, zero non-radiative heating source~\eqref{gasgrain}, arbitrary incident radiation. 
  In this case the mean intensity in the media is equal to the incident one and temperature may be derived from 
  \eqref{EEBFinal} and it is the same for every position in media. Tests show an equality of temperatures which 
  are derived by our RT code and by independent numerical solution of the energy balance equation \eqref{EEBFinal}.
  \item Arbitrary media (optically thick, non-gray opacities) with the non-diluted Planck incident radiation. 
  In this case the media becomes isothermal with temperature of radiation. Test show an equality of temperatures 
  and radiation field derived from analytical and numerical solutions.
  \item Optically thick media with gray opacities, arbitrary incident radiation and non-radiative dust heating 
  function which is proportional to dust density and has no other $z$-dependence. In this case the system of 
  Equations~\eqref{MainEq}--\eqref{EEBFinal} narrow down to the second order linear ODE and may be solved 
  analytically. A comparison of the analytical and numerical solutions is presented in Figure~\ref{RTbenchFig}. 
  The mean intensity parabolic profile is shown in the inset graph as well.
\end{itemize}

%---------------------------------------------------------------
\begin{figure}[ht]
\begin{center}
$
\begin{array}{cc}
	\includegraphics[ scale=.34, angle=270]{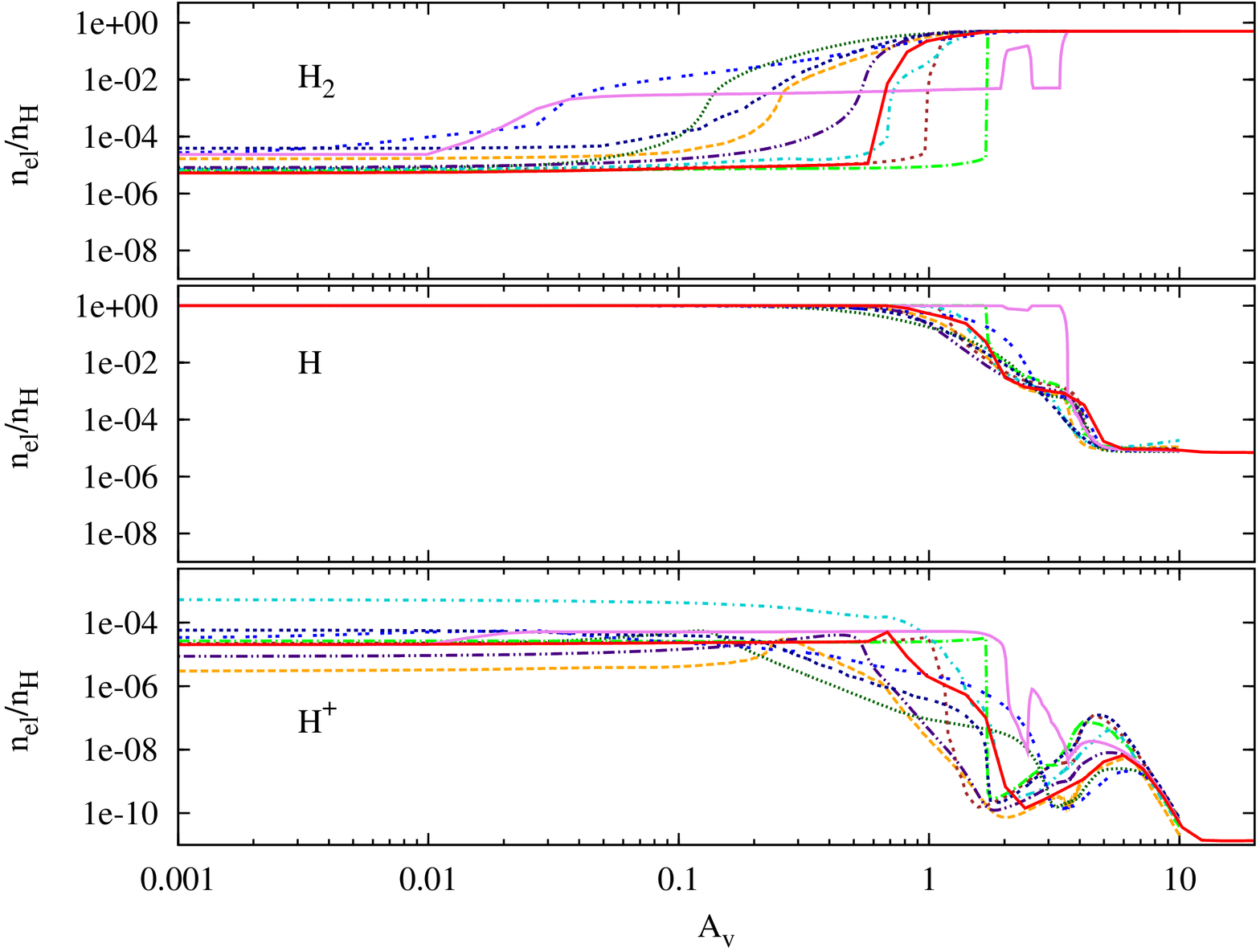} & 
	\includegraphics[ scale=.34, angle=270]{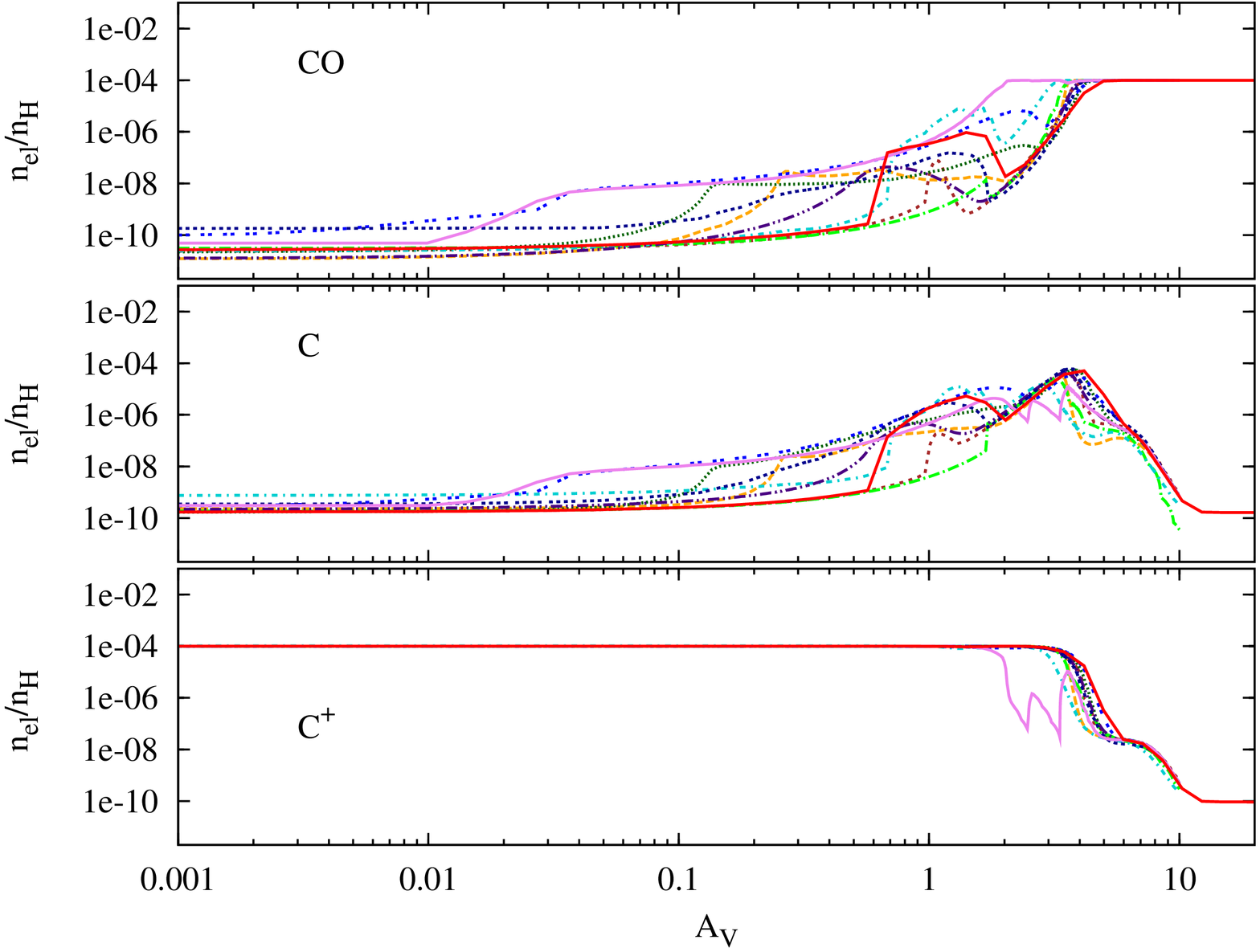} 
\end{array}
$
\end{center}
\vspace{-.5cm}
 \caption{\textit{Left panel.} Comparison of H$_2$, H, H$^+$ density profiles calculated by our code (solid red line) with post-benchmark results from \cite{Rollig2007} (code markings are the same as in Figure~\ref{bench_T}) for benchmark model V4 ($n=10^{5.5} \cmc$, $\chi=10^{5}$). \textit{Right panel.}  The same for C, C$^+$, CO density profiles.} 
\label{bench_abund}
 \end{figure}
%---------------------------------------------------------------

%---------------------------------------------------------------

 \begin{figure}[!ht]
 $\begin{array}{cc}
	\includegraphics[ scale=.35, angle=270]{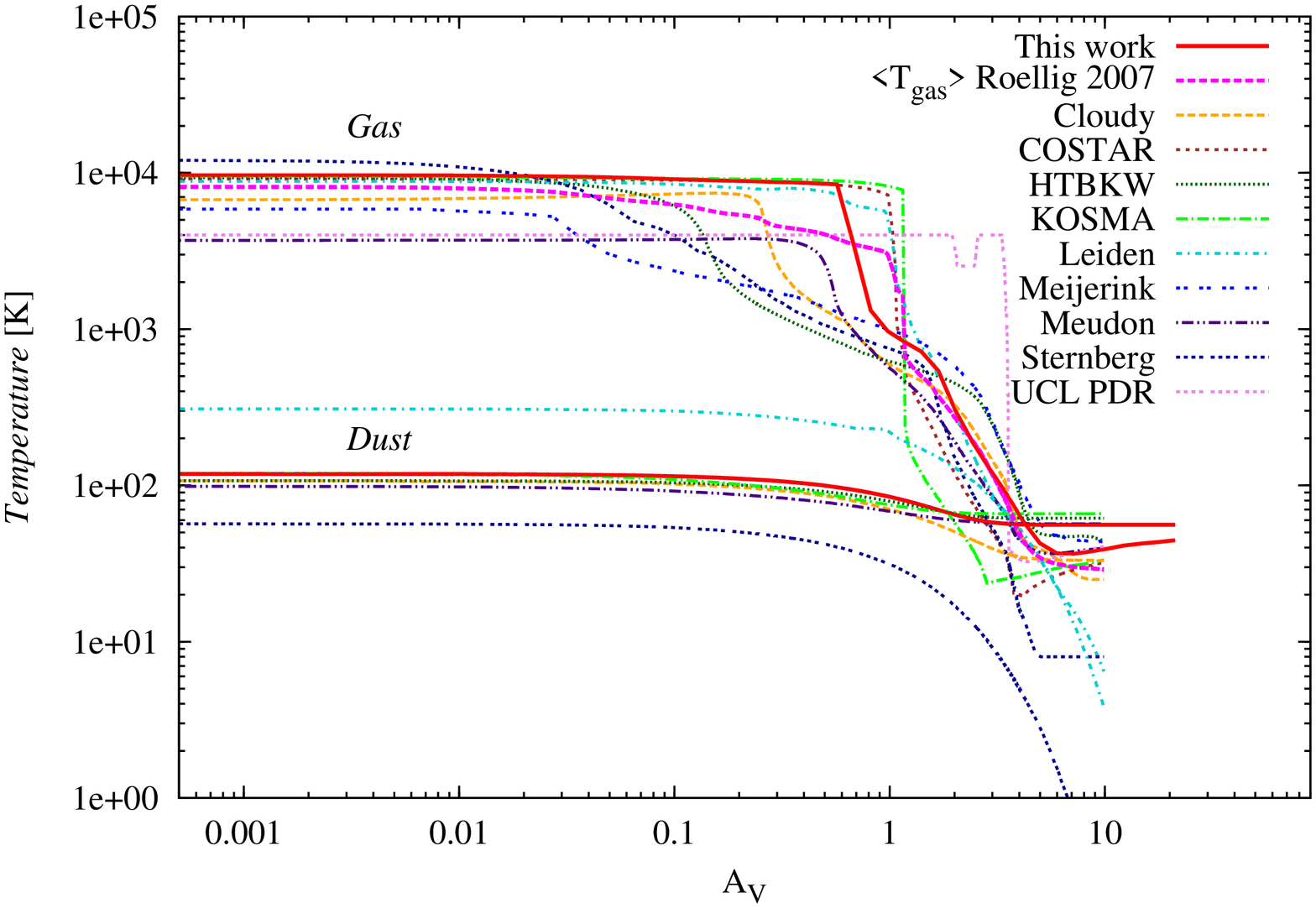} & 
	\includegraphics[ scale=.35, angle=270]{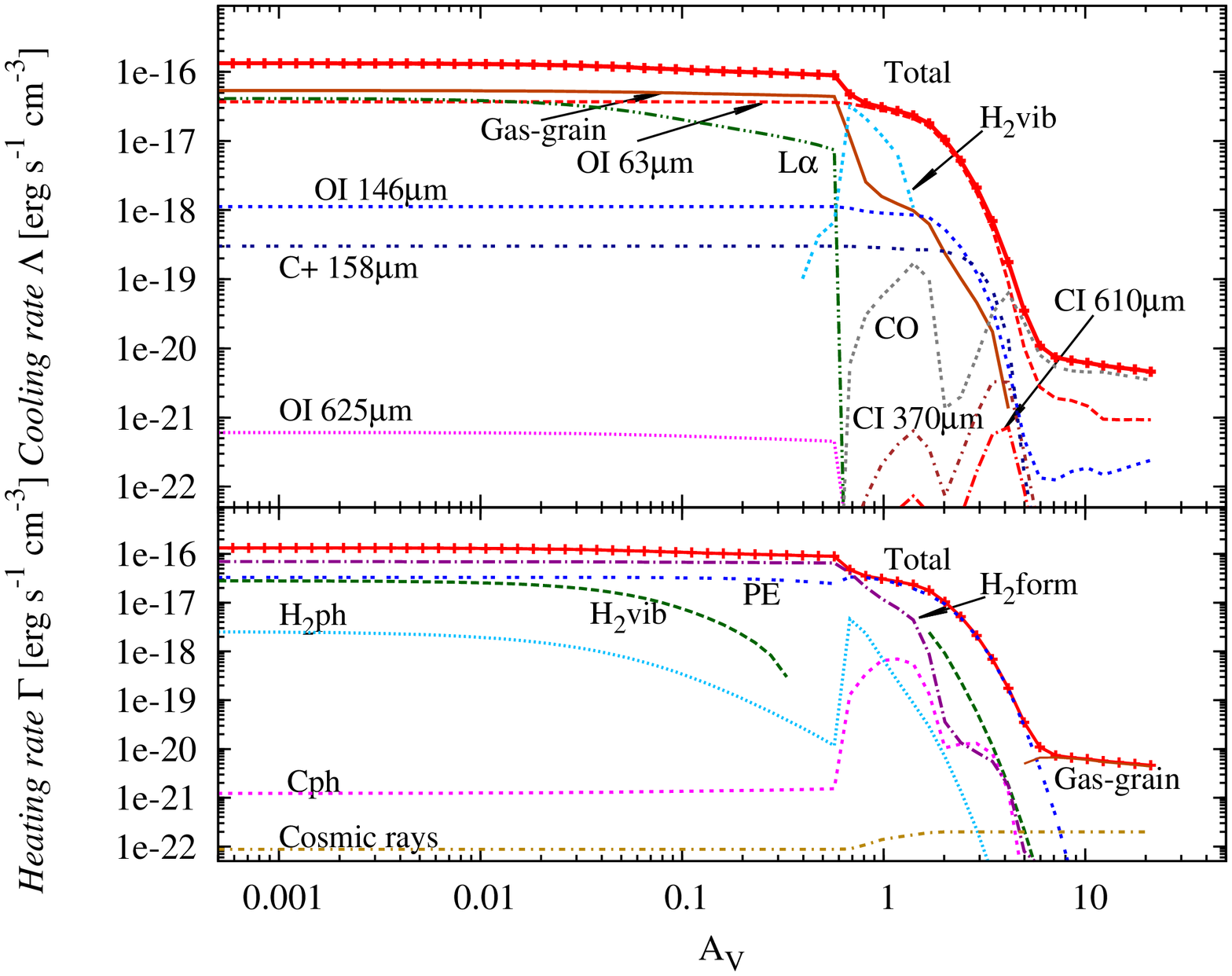} 
\end{array}$
 \caption{\textit{Left panel.} Comparison of gas temperature calculated by our code (red solid curve) with 
 post-benchmark results from different PDR codes from \cite{Rollig2007}. Thick magenta dashed line marks 
 average gas temperature from the benchmarking data.
 \textit{Right panel. } Main heating and cooling processes included in our code for the benchmark model V4.}
\label{bench_T} 
 \end{figure}
%---------------------------------------------------------------

%-------------------------------------

The surface of the disk is a photodissociation region (PDR) controlled by UV radiation from a star and interstellar radiation field. Therefore, we perform benchmarking of the thermal balance in our code as proposed in the PDR code comparison study \citep{Rollig2007}. For benchmarking purposes we use a reduced chemical network restricted to the most abundant elements (H, He, O, C, e$^-$) and 31 species(Table 4 in \citealp{Rollig2007}). The reaction rates are taken from the UMIST99 database with some corrections from A. Sternberg.  \HH dissociation rate is $5\times 10^{-18} \chi/10\, \rm s^{-1}$. Cosmic ray H ionization rate is $\zeta = 5 \times 10^{-17}\, \rm s^{-1}$. For more details of benchmark test we refer to \cite{Rollig2007}. 

All benchmark models assume plane-parallel, semi-infinite geometry of clouds of total constant hydrogen density of $10^3$ and $10^{5.5}\cmc$. The values of the standard far UV field were taken as $\chi=10$ and $10^5$ times the \cite{Draine1978} field. There are two sets of benchmark models: four with fixed dust and gas temperatures of 30 and 50~K, respectively, and the other set of four models with the gas temperature resulting from thermal balance. The first set of models with fixed temperature aims at testing main ingredients of the thermal balance: solutions of chemistry and statistical equilibrium equations for level populations of main coolants, while the second set examines solution of thermal balance. Here we present results of benchmark tests for both kinds of models with density $n^{\rm tot}_{H}=10^{5.5}\cmc$ and far UV field strength $\chi=10^5$ (models F4 and V4 in \cite{Rollig2007}).

The left panel of Figure~\ref{bench_abund} shows comparison of our calculations with post-benchmark results for the H/{\HH} 
transition zone typical for PDR environment. Right panel of Figure~\ref{bench_abund} shows the C$^+$/CO/C transition zone. 

Main heating and cooling rates included in benchmarking are shown in the left panel of Figure~\ref{bench_T}. 
Gas-grain cooling and [OI] 63\mum line are the dominant cooling processes for $A_V<0.5$. CO lines dominate 
cooling at high attenuated regions. Our line cooling rates show remarkable agreement with data from 
\citet{Rollig2007} for dominant  cooling processes: [CII] 158\mum, [OI] 63, 145\mum, [CI] 370, 610\mum lines.
Comparison of our model results for gas temperature in the slab with other PDR codes is shown in Figure~\ref{bench_T}. 
At small $A_V$ the gas temperature is much higher than the dust temperature due to photoelectric heating and agrees well with other PDR codes.